\documentclass{siamltex}

\usepackage[figuresright]{rotating}

\usepackage{amsmath}

\usepackage{graphics}
\usepackage{algorithm}
\usepackage{algorithmic}
\usepackage{amssymb}
\usepackage{listings}

\title{Compressed Multi-Row Storage Format for Sparse Matrices on Graphics Processing Units}

\author{Zbigniew Koza$^*$, Maciej Matyka$^*$, Sebastian Szkoda\thanks{Faculty of Physics and Astronomy, University of Wroc{\l}aw,
         Wroc{\l}aw, Poland (zkoza@ift.uni.wroc.pl).}
         \and {\L}ukasz Miros{\L}aw\thanks{Institute of Informatics, Wroc{\l}aw University of Technology, Wroc{\l}aw, Poland and
         Vratis Ltd., Wroc{\l}aw, Poland}}

\begin{document}

\maketitle

\begin{abstract}
A new format for storing sparse matrices is proposed for efficient sparse matrix-vector (SpMV) product calculation
on modern graphics processing units (GPUs).
This format extends the standard compressed row storage (CRS) format and can be quickly converted to and from  it.
Computational performance of two SpMV kernels for the new format is determined for over 130 sparse
matrices on Fermi-class and Kepler-class GPUs and compared with that of five existing generic algorithms and
industrial  implementations, including Nvidia cuSparse CSR and HYB kernels.
We found the speedup of up to $\approx 60\%$ over the best of the five alternative kernels.

\end{abstract}



%%%%%%%%%%%%%%%%%%%%%%%%%%%%%%%%%%%%%%%%%%%%%%%%%%%%%%%%%%%%
%%%%%%%%%%%%%%%%%%%%%%%%%%%%%%%%%%%%%%%%%%%%%%%%%%%%%%%%%%%%
%%%%%%%%%%%%%%%%%%%%%%% SECTION %%%%%%%%%%%%%%%%%%%%%%%%%%%%
%%%%%%%%%%%%%%%%%%%%%%%%%%%%%%%%%%%%%%%%%%%%%%%%%%%%%%%%%%%%
%%%%%%%%%%%%%%%%%%%%%%%%%%%%%%%%%%%%%%%%%%%%%%%%%%%%%%%%%%%%

\section{Introduction}\label{sec:introduction}

The sparse matrix-vector (SpMV) multiplication is one of the most important  kernels in scientific computing
with numerous applications ranging from sparse linear solvers to the PageRank
algorithm used by Google in its Web search engine.
However, its high memory bandwidth
requirements combined with the poor data locality exhibited by typical sparse
matrices result in poor performance on general purpose processors which usually
attain only a small fraction of their peak  performance in this kernel.
The literature devoted to SpMV optimization techniques on traditional,
cache-based processor designs is ample (see \cite{Vuduc:Aut2003} for an extensive
overview), and currently one of the major issues is how to exploit new features
available in multicore hardware. Several SpMV optimization strategies were
already proposed for various designs of chip multiprocessors, including
multicore central processing units (CPUs) from Intel and AMD
\cite{Williams:Opt2007}, single- and dual-socket STI Cell
\cite{Williams:Opt2007}, Sun UltraSPARC T2  \cite{Williams:Opt2007}, field
programmable gate arrays (\mbox{FPGAs}) \cite{DuBois:AnI2008,DuBois:Spa2008},
Intel Xeon Phi coprocessors \cite{Liu:2013,Saule:arxiv:2013},
and graphics processing units (GPUs)
\cite{Bell08,Bell09,Baskaran2009,Dziekonski11,Feng2011,Maggioni13,Monakov10,Mukunoki2013,Vazquez2009,Vazquez2010,Yang11}.

In just a few years GPUs have evolved from fixed-pipeline application-specific integrated
circuits
into highly programmable, versatile computing devices
with peak computational performance matching that of the most powerful supercomputers
of only a decade ago.
These devices can be programmed in high-level
programming languages, e.g.\ Nvidia's C++ for CUDA (proprietary) or OpenCL (open standard).
The major problem in their usage is software: any new hardware architecture will
succeed only if appropriate software can be developed to exploit the parallelism
in the hardware efficiently.
However, the current multicore architectures are so diverse and are subject to so frequent changes
that to exploit their potential the applications must be highly specialized
and use architecture-specific optimization strategies. Moreover, massively parallel architectures,
like  the one utilized in modern GPUs, require to use a
new programming paradigm,
which in turn requires radical rethinking of how numerical computations should be performed.
Since the vast majority of the existing scientific software adheres to  ``old''
programming paradigms, and since development of this software took millions of man-hours,
perhaps the best way to introduce new concepts is by concentrating on the most important kernels
and showing their usability in the ``old'' environment.

In this paper we present an efficient SpMV algorithm optimized for
processors with programmable on-chip shared memory and examine its implementation for Nvidia's CUDA-enabled GPUs.
The algorithm is based on a new sparse matrix storage format, CMRS, designed as an extension of a
popular compressed row storage (CRS, also known as compressed sparse row, CSR) format. It is characterized by
a small memory footprint and small conversion times to other storage formats, which should facilitate
its adoption in existing applications.
It also turns out to be among the fastest SpMV algorithms available for GPUs.
Moreover, in contrast to many recent studies on GPU SpMV,
which were based on small sets of sparse matrices, which in turn almost reduced the papers to case studies,
here we use a far larger set from the University of Florida Sparse Matrix Collection (UF SMC) \cite{Florida}.
This enabled us to make some general statements not only about the absolute performance of our CMRS-based implementation,
but also about relative performance of several other alternative solutions for two generations of
GPU architectures.

%%%%%%%%%%%%%%%%%%%%%%%%%%%%%%%%%%%%%%%%%%%%%%%%%%%%%%%%%%%%
%%%%%%%%%%%%%%%%%%%%%%%%%%%%%%%%%%%%%%%%%%%%%%%%%%%%%%%%%%%%
%%%%%%%%%%%%%%%%%%%%%%% SECTION %%%%%%%%%%%%%%%%%%%%%%%%%%%%
%%%%%%%%%%%%%%%%%%%%%%%%%%%%%%%%%%%%%%%%%%%%%%%%%%%%%%%%%%%%
%%%%%%%%%%%%%%%%%%%%%%%%%%%%%%%%%%%%%%%%%%%%%%%%%%%%%%%%%%%%

\section{Problem statement\label{sec:problem:statement}}
\subsection{SpMV multiplication}

The aim of the SpMV multiplication algorithm is to calculate the
product $\mathbf{y}=\hat{A}\mathbf{x}$, where $\hat{A}$ is a large sparse matrix
and $\mathbf{x}$, $\mathbf{y}$ are dense vectors.
Typical matrices involved in the SpMV product  have thousands or even millions of rows and columns,
but the average number of nonzero elements per row rarely exceeds 100.
The most
interesting---and difficult---matrices are those whose distribution of nonzero elements appears to be unpredictable.

Nonzero elements of $\hat A$ are usually stored in an auxiliary array,
and additional information is needed to uniquely map the values in the array to their locations in $\hat A$.
The way this information is stored is called a sparse matrix format.
Calculation of a sparse matrix-vector product essentially reduces to many ``multiply and add'' operations,
which in modern GPUs are implemented as a single fused multiple-add (FMA) instruction. Since
SpMV multiplication  involves several memory accesses per
arithmetic instruction, the SpMV kernel is inherently memory-bound.
For example, a server-class Tesla K20X GPU can perform $\approx 6.5 \times 10^{11}$ FMA operations per second
and can access its main memory at $\approx 2.5 \times 10^{11}$ B/s,
which yields approximately 2.5 operations per byte. For the SpMV kernel this
sets the upper bound for the processor computational efficiency
to  $\approx 1/80$ of its peak theoretical value.
Although this value can be increased by using fast on-chip caches,
other factors, like additional memory transactions necessary to read sparse matrix
format data or reduced off-chip memory throughput due to poor data locality
can decrease it to even smaller values.
The main challenge
is thus how
to exploit and balance all the performance-related features available in  hardware,
focusing on the utilization of the memory.

%%%%%%%%%%%%%%%%%%%%%%%%%%%%%%%%%%%%%%%%%%%%%%%%%%%%%%%%%%%%
%%%%%%%%%%%%%%%%%%%%% SUB-SECTION %%%%%%%%%%%%%%%%%%%%%%%%%%
%%%%%%%%%%%%%%%%%%%%%%%%%%%%%%%%%%%%%%%%%%%%%%%%%%%%%%%%%%%%

\subsection{GPU architecture and the CUDA programming model}
The architecture of modern GPUs is a massively parallel design which excels
in computing-intensive stream data processing.
Here we briefly discuss the main properties of the ``Fermi'' (2010) and ''Kepler'' (2012) GPU architectures
from Nvidia \cite{CUDA55}.

A GPU contains a number of units called multiprocessors,
each one containing a set of relatively simple computing cores called CUDA cores.
From the programmer's perspective, multiprocessors are essentially independent single-instruction multiple data (SIMD)
devices in which groups of 32 CUDA threads, called warps, execute the same instruction on multiple data simultaneously.
Multiprocessors are connected to high bandwidth (up to $\approx$ 290~GB/s), high latency ($\approx$ 800 clock
cycles), limited size ($\leq $12~GB) external dynamic random access memory through a coherent L2 cache (up to 1.5~MB).
Each multiprocessor has an L1 cache, a read-only texture cache, a 48-KB read-only cache (Kepler K20 and K40 GPUs only)
and a constant cache (8~KB).
Beside these hardware-managed caches, multiprocessors contain also several fast on-chip memories managed in software:
so called shared memory (up to 48 KB) and registers. The shared memory
is shared by all threads belonging to well-defined groups of warps (called blocks) executing on the same multiprocessor.
The sizes of these resources are available and, to some extent, configurable at run-time.

Various memories available in GPUs differ not only in their size and speed, but also in latencies.
For example, the latency of registers
is $\approx 20$ clock cycles, whereas the latency of the global memory accesses can be as high
as 800 clock cycles. To hide such high latencies, each multiprocessor loads into its registers the states of up
to 1536  (Fermi) or 2048 (Kepler) threads and attempts to execute the warp that has all operands ready for execution.
This leads to massive parallelism with thousands of threads being processed on-chip simultaneously.
For this approach to be efficient, the occupancy, defined as the ratio of the number of resident threads to
the maximum number of resident threads, must be sufficiently high.
Another factor crucial for GPU efficiency is the memory access pattern.
For example,
the condition for the global memory to be utilized at full speed is that
all threads in a warp should access contiguous, 128-byte aligned locations.

CUDA is an abstract general purpose parallel computing architecture, programming model, and programming environment
designed for Nvidia's GPUs \cite{Farber2011}.
It is based on a few key concepts such as groups of threads (arranged hierarchically in warps, blocks and a grid),
shared memories and barrier synchronization. These concepts are exposed to the programmer
through a minimal set of extensions to a high-level programming language  (C or C++).
Warps within a block of threads can be executed in any order; similarly, each block of threads can be run on
any of the available multiprocessors in an arbitrary order, sequentially or in parallel.

%%%%%%%%%%%%%%%%%%%%%%%%%%%%%%%%%%%%%%%%%%%%%%%%%%%%%%%%%%%%
%%%%%%%%%%%%%%%%%%%%% SUB-SECTION %%%%%%%%%%%%%%%%%%%%%%%%%%
%%%%%%%%%%%%%%%%%%%%%%%%%%%%%%%%%%%%%%%%%%%%%%%%%%%%%%%%%%%%

\subsection{Existing matrix formats for SpMV multiplication\label{sub:existing-formats}}

The simplest sparse matrix format is the coordinate (COO) format, in which
the information about the row index, column index, and the
value of each non-zero matrix element is stored in three one-dimensional arrays,
\texttt{RowInd}, \texttt{ColInd}, and \texttt{Val}, respectively.
As an example, consider  a $5\times5$ matrix:

\begin{equation}
 \hat M = \left[
 \begin{array}{ccccc}
   1 & 0 & 0 & 2 & 0 \\
   0 & 3 & 0 & 0 & 4 \\
   0 & 0 & 5 & 0 & 6 \\
   0 & 0 & 7 & 8 & 9 \\
   0 & 0 & 0 & 0 & 10
 \end{array}
\right].
\end{equation}
Its COO representation (with zero-based indexing) reads
\begin{eqnarray*}
\texttt{Val}    &=& \left[\; 1 \;\; 2 \;\; 3 \;\; 4 \;\; 5 \;\; 6 \;\; 7 \;\; 8 \;\; 9 \;\; 10 \; \right], \\
\texttt{ColInd} &=& \left[\; 0 \;\; 3 \;\; 1 \;\; 4 \;\; 2 \;\; 4 \;\; 2 \;\; 3 \;\; 4 \;\; 4  \; \right], \\
\texttt{RowInd} &=& \left[\; 0 \;\; 0 \;\; 1 \;\; 1 \;\; 2 \;\; 2 \;\; 3 \;\; 3 \;\; 3 \;\; 4  \; \right].
\end{eqnarray*}
To complete the matrix definition,
one also needs to supply three integers:
the number of matrix rows (\texttt{rows}), columns (\texttt{cols})
and non-zero elements (\texttt{nnz}).

In the above example the row-major ordering was used,
i.e., the matrix index arrays were first sorted by row indices and then by column indices.
In such a case array \texttt{RowInd} will typically contain sequences of many identical entries.
This property is utilized in the CRS format to reduce the memory footprint
by replacing array \texttt{RowInd} with a shorter array
\texttt{RowPtr}. In the most general case this array is defined by the requirement that
\texttt{RowPtr[j+1] - \texttt{RowPtr}[j]}
 be equal to the number of non-zero elements in the $j$-th row ($j=0,\ldots,\mathtt{rows}-1$).
If the matrix contains no empty rows, \texttt{RowPtr}[j] gives the index into \texttt{Val}
corresponding to the first non-zero element in the $j$-th matrix row.
Array \texttt{RowPtr} has exactly \texttt{rows}$ + 1$ elements and \texttt{RowPtr[rows]} = \texttt{nnz}.
Thus, the CRS representation of $\hat M$ reads
\begin{eqnarray*}
\texttt{Val}    &=& \left[\; 1 \;\; 2 \;\; 3 \;\; 4 \;\; 5 \;\; 6 \;\; 7 \;\; 8 \;\; 9 \;\; 10 \;\right], \\
\texttt{ColInd} &=& \left[\; 0 \;\; 3 \;\; 1 \;\; 4 \;\; 2 \;\; 4 \;\; 2 \;\; 3 \;\; 4 \;\; 4 \;\right], \\
\texttt{RowPtr} &=& \left[\; 0 \;\; 2 \;\; 4 \;\; 6 \;\; 9 \;\; 10 \;\right].
\end{eqnarray*}
Note that arrays \texttt{Val} and \texttt{ColInd} are the same as in the COO format.

Let $k$ be the maximum number of non-zero elements per row.
In the ELLPACK/ITPACK (ELL) format an $n\times m$ sparse matrix is represented by two $n\times k$ dense arrays,
\texttt{Val} and \texttt{ColInd}. Array \texttt{Val} is constructed from the original matrix by removing all zeros, while
 \texttt{ColInd} holds column indices into \texttt{Val}.
The rows with less than $k$ non-zero elements are padded in \texttt{Val} and \texttt{ColInd} arrays with
$0$ and $-1$, respectively.
The ELL representation of $\hat M$ is thus:
\begin{equation*}
 \texttt{Val} = \left[
  \begin{array}{ccc}
     1 & 2 & 0 \\
     3 & 4 & 0 \\
     5 & 6 & 0 \\
     7 & 8 & 9 \\
     10& 0 & 0 \\
  \end{array}
 \right],
\hspace{0.025\textwidth}
 \texttt{ColInd} = \left[
  \begin{array}{rrr}
     0 &  3 & -1 \\
     1 &  4 & -1 \\
     2 &  4 & -1  \\
     2 & 3 & 4 \\
     4 & -1 & -1 \\
  \end{array}
 \right].
\end{equation*}

While the ELL format belongs to the most efficient sparse matrix formats for vector architectures, it
may involve a costly storage overhead.
Several attempts have been made to modify this format so as to extend its practical usability for general sparse matrices.
One such attempt is the hybrid (HYB) format \cite{Bell08,Bell09},
which is a combination of the ELL and COO formats. Another idea is to divide the matrix into several slices,
each represented separately in the ELL format, and/or use some kind of matrix transformation,
e.g.\ permutation of rows \cite{Monakov10,Dziekonski11}, to reduce padding.

%%%%%%%%%%%%%%%%%%%%%%%%%%%%%%%%%%%%%%%%%%%%%%%%%%%%%%%%%%%%
%%%%%%%%%%%%%%%%%%%%% SUB-SECTION %%%%%%%%%%%%%%%%%%%%%%%%%%
%%%%%%%%%%%%%%%%%%%%%%%%%%%%%%%%%%%%%%%%%%%%%%%%%%%%%%%%%%%%

\subsection{Existing GPU implementations\label{sub:existing-implementations}}

One of the first efficient implementations of SpMV on the GPU architecture were proposed
by Bell and Garland \cite{Bell08,Bell09}.
They implemented SpMV kernels for several sparse matrix formats, including COO, ELL, and HYB.
In addition, two SpMV kernels for the CRS format were provided: scalar and vector.
The scalar kernel assigns one thread
per matrix row, which results in  non-coalesced access to memory and poor performance.
The vector kernel assigns a 32-thread warp to each row---while this ensures
contiguous access to the memory, it leads to a large
bandwidth waste whenever a row size is much smaller than the warp size.
As for  COO, the tests showed that it is
not flexible enough to handle unstructured matrices efficiently.
The ELL format is often the fastest, but fails whenever row sizes vary significantly,
as it leads to a large memory overhead. This problem
was addressed in  the HYB format, in which the matrix
is partitioned into a regular part, stored in  ELL, and an irregular part,
stored in COO \cite{Bell09,CUDA55}. The partitioning of a general matrix is
a rather complex operation which
requires building a histogram of the row sizes to find the balance between the
potential storage overhead of ELL and the computational inefficiency of COO.
The authors recommended HYB as the fastest format for a
broad selection of unstructured matrices.

Bell's and Garland's SpMV kernels served as building blocks for the CUSP \cite{CUSP} library.
To improve coalescing of matrix data accesses  for matrices in the CRS representation,
CUSP can virtually divide each warp into 2, 4, 8 or 16 smaller parts
and assign them to different rows. Mukunoki and
Takahashi used the same idea to optimize their CRS kernel for the Kepler GPU architecture \cite{Mukunoki2013}.
Baskaran and Bordawekar \cite{Baskaran2009} proposed a few other optimization techniques based
on exploiting synchronization-free parallelism and optimized off-chip memory access.
Another direction of research on improving the efficiency of the SpMV kernel on GPUs
focuses on various extensions and modifications of  the ELL or CRS
formats. This resulted in the development of the ELL-R \cite{Vazquez2009},
sliced-ELL \cite{Monakov10}, ELLR-T \cite{Vazquez2010,Vazquez2012}, and
Sliced ELLR-T \cite{Dziekonski11}
formats, tiling and composite storage \cite{Yang11}, as well as the CRS-T \cite{Yoshizawa2012}
and CSR SIC \cite{Feng2011} formats.

%%%%%%%%%%%%%%%%%%%%%%%%%%%%%%%%%%%%%%%%%%%%%%%%%%%%%%%%%%%%
%%%%%%%%%%%%%%%%%%%%%%%%%%%%%%%%%%%%%%%%%%%%%%%%%%%%%%%%%%%%
%%%%%%%%%%%%%%%%%%%%%%% SECTION %%%%%%%%%%%%%%%%%%%%%%%%%%%%
%%%%%%%%%%%%%%%%%%%%%%%%%%%%%%%%%%%%%%%%%%%%%%%%%%%%%%%%%%%%
%%%%%%%%%%%%%%%%%%%%%%%%%%%%%%%%%%%%%%%%%%%%%%%%%%%%%%%%%%%%

\section{Compressed Multi-Row Sparse Format\label{sec:CMRS}}

Efficiency of existing GPU implementations of the SpMV product is often significantly
better if an ELL-based format, e.g.\ HYB, is used instead of CRS.
The main reason for this is that GPUs are SIMD-like machines with relatively
wide SIMD units, often far wider than the average row length.
Since efficient utilization of the  CRS format requires the matrix elements to be accessed row by row,
processing short rows in long SIMD-like units leads to  wasting of the computational capability of the device.
Therefore, our main idea is to process a sparse matrix in chunks larger than individual rows,
at the same time preserving the overall structure of the matrix representation typical of the CRS format.
A group of rows processed by an individual SIMD unit shall be called `strip', and the number of
rows in a strip shall be called `strip height' and denoted \texttt{height}.
The number of strips, \texttt{strips} is
thus equal to  \texttt{ceil(rows/height)}.

The new sparse matrix format, which we call compressed multi-row storage (CMRS) format, comprises  one integer
parameter \texttt{height} and four arrays: data array  \texttt{Val} and three auxiliary integer arrays,
\texttt{ColInd}, \texttt{StripPtr}, and \texttt{RowInStrip}.
Arrays  \texttt{Val} and  \texttt{ColInd} are the same as in the CRS format.
Array \texttt{StripPtr} is a generalization of CRS array \texttt{RowPtr} and
is defined by the requirement that \texttt{StripPtr[j+1] - \texttt{StripPtr}[j]} be equal
to the number of non-zero elements in the $j$-th strip ($j=0,\ldots,\mathtt{strips}-1$).
If the sparse matrix contains no empty strips, \texttt{StripPtr}[j] gives the index into \texttt{Val}
corresponding to the first non-zero element in the $j$-th strip.
Finally, array \texttt{RowInStrip}, of length \texttt{nnz}, holds the row numbers within individual strips.

Assume $\mathtt{height} = 2$. Then the CMRS representation of $\hat M$ reads:
\begin{eqnarray*}
\texttt{Val}    &=& \left[\; 1 \;\; 2 \;\; 3 \;\; 4 \;\; 5 \;\; 6 \;\; 7 \;\; 8 \;\; 9 \;\; 10 \;\right], \\
\texttt{ColInd} &=& \left[\; 0 \;\; 3 \;\; 1 \;\; 4 \;\; 2 \;\; 4 \;\; 2 \;\; 3 \;\; 4 \;\; 4 \;\right], \\
\texttt{StripPtr} &=& \left[\; 0 \;\; 4 \;\; 9 \;\; 10 \;\right],\\
\texttt{RowInStrip} &=& \left[\; 0 \;\; 0 \;\; 1 \;\; 1 \;\; 0 \;\; 0 \;\; 1 \;\; 1 \;\; 1 \;\; 0 \;\right].
\end{eqnarray*}
Note that the conversion between the CRS and CMRS formats is trivial and easy to parallelize.
In particular, $\mathtt{StripPtr[j]} =  \mathtt{RowPtr[j*height]}$ for $j < \mathtt{strips}$ and
$\mathtt{StripPtr[strips]} = \mathtt{nnz}$, whereas \texttt{RowInStrip[k]} is the remainder
of the row number divided by \texttt{height}.
It is also clear that both formats are equivalent if $\mathtt{height} = 1$, hence CMRS can be regarded as an extension
of the CRS format.

The idea that a warp could process a  group of adjacent matrix rows was already exploited in
Refs.~\cite{CUSP,Feng2011,Yoshizawa2012,Mukunoki2013}.
They all used a static mapping of warp threads to the rows, which is
inefficient if the lengths of adjacent rows vary significantly. In particular,
the CSR SIC format \cite{Feng2011} interleaves several matrix rows to form a new SIC row.
This, however, requires padding shorter rows with explicit zeroes.
As this could easily lead to  prohibitive memory overhead, the rows must be first reordered
according to their lengths, then combined into larger SIC rows,
and these are then combined into a few large segments processed by separate GPU kernels.
The CMRS format solves these problems by dynamically assigning threads to rows through
the \texttt{RowInStrip} array. An  efficient CMRS-based SpMV kernel requires neither zero-padding
nor row reordering and can be implemented as a single GPU kernel.

%%%%%%%%%%%%%%%%%%%%%%%%%%%%%%%%%%%%%%%%%%%%%%%%%%%%%%%%%%%%
%%%%%%%%%%%%%%%%%%%%%%%%%%%%%%%%%%%%%%%%%%%%%%%%%%%%%%%%%%%%
%%%%%%%%%%%%%%%%%%%%%%% SECTION %%%%%%%%%%%%%%%%%%%%%%%%%%%%
%%%%%%%%%%%%%%%%%%%%%%%%%%%%%%%%%%%%%%%%%%%%%%%%%%%%%%%%%%%%
%%%%%%%%%%%%%%%%%%%%%%%%%%%%%%%%%%%%%%%%%%%%%%%%%%%%%%%%%%%%

\section{Implementation\label{sec:implementation}}

Our GPU implementation of the SpMV product for matrices in the CMRS format is based on the vector kernel by
Garland and Bell \cite{Bell09}, with rows replaced by strips. Each SIMD unit, or warp made of $\mathtt{W\_SIZE} = 32$ threads,
is assigned a strip to process.
The method of doing the SpMV product in a strip is presented in Algorithm~\ref{alg:CMRS} and explained below.
Note that we used 28 bits of \texttt{ColInd[j]} to
store a column index and the remaining 4 bits (denoted as \texttt{CMRS\_BITS})
to store the corresponding value of \texttt{RowInStrip}. This made array
\texttt{RowInStrip} superfluous and explains `uncompression' steps in Algorithm~\ref{alg:CMRS}.

\begin{algorithm}
\caption{Processing of a strip in the SpMV kernel, $\mathbf{y} = \hat{A} \cdot \mathbf{x}$, for $\hat{A}$ in the CMRS format.
This algorithm is executed in parallel by \texttt{W\_SIZE} (``warp size'') threads
identified by $\texttt{thread\_lane} \in \{0,1,\ldots,\mathtt{W\_SIZE}-1\}$ and making up a warp.
Parameter \texttt{strip\_id} identifies the strip in the matrix
and  \texttt{height} is the number of rows making up a strip.
Suggested value of \texttt{CMRS\_BITS} is 4. \label{alg:CMRS}}
\begin{algorithmic}
\REQUIRE \texttt{Val}, \texttt{ColInd},  \texttt{StripPtr}, \texttt{height},
$\mathbf{x}$,  $\mathbf{y}$, $0\le\mathtt{strip\_id}<\mathtt{strips}$
\STATE
 \STATE  $\mathtt{buf}_{i,j} \gets 0$  for all $i,j$
\STATE $M \gets 2^\mathtt{CMRS\_BITS}$
\STATE $\mathtt{strip\_start} \gets \mathtt{StripPtr}_\mathtt{strip\_id}$
\STATE $\mathtt{strip\_end} \gets \mathtt{StripPtr}_{\mathtt{strip\_id}+1}$
\STATE \COMMENT{$j$ is the current  index into \texttt{Val} and \texttt{ColInd}}
\STATE $j \gets \mathtt{strip\_start} +  \mathtt{thread\_lane}$
\WHILE {$j < \mathtt{strip\_end}$}
   \STATE \COMMENT{Load compressed values into register}
   \STATE $c \gets \mathtt{ColInd}_j$
   \STATE  \COMMENT{Uncompress the value of \texttt{RowInStrip}$_j$}
   \STATE $r \gets c \mod M$
   \STATE  \COMMENT{Uncompress the value of \texttt{ColInd}$_j$}
   \STATE $c \gets \lfloor c / M \rfloor$
   \STATE  \COMMENT{Threads update partial sums}
   \STATE $\mathtt{buf}_{\mathtt{thread\_lane},r}  \gets \mathtt{buf}_{\mathtt{thread\_lane},r} + x_c \cdot \mathtt{Val}_j$
   \STATE $j \gets j + \mathtt{W\_SIZE}$
\ENDWHILE
   \STATE
   \STATE \COMMENT{Parallel reduction of partial sums in rows}
   \STATE $\mathtt{buf}_{0,r} \gets  \sum_{i=0}^{\mathtt{W\_SIZE}-1} \mathtt{buf}_{i,r}$,
$r=0,\ldots,\mathtt{height}-1$
   \STATE
   \STATE $\mathtt{row} \gets \mathtt{strip\_id} \cdot \mathtt{height} + \mathtt{thread\_lane}$
   \STATE \COMMENT{\texttt{height} elements of \texttt{buf}$_0$ contain row sums}
   \IF {$\mathtt{thread\_lane} < \mathtt{height}$ \AND $\mathtt{row} < \mathtt{num\_rows}$}
      \STATE $y_\mathtt{row} \gets \mathtt{buf}_\mathtt{0,thread\_lane}$
   \ENDIF
\end{algorithmic}
\end{algorithm}

This algorithm is assumed to be executed in parallel by all threads in a warp. Implicit synchronization
of the threads forming a warp is assumed. All auxiliary buffers and temporary variables are local to a warp, so no explicit
synchronization of different warps is necessary, which allows for massively parallel processing of strips.
The values of matrix elements and the corresponding column indices are read in parallel directly from arrays
\texttt{Val} and \texttt{ColInd}, whereas row indices are assembled from the information held in arrays
\texttt{StripPtr} and \texttt{RowInStrip}.  For sufficiently long strips these memory operations are coalesced to a
high degree ($j$ runs through consecutive matrix elements), and hence are very fast.
Then the necessary elements of the input vector are fetched from the memory.
This is the most sensitive part of each parallel SpMV implementation, as the vector elements required by a SIMD unit are
often stored in memory locations scattered almost randomly in the memory, and hence their parallel
processing is very problematic. Individual products of the  matrix by vector elements are
computed and stored in a buffer \texttt{buf} allocated in the fast on-chip shared memory.
The buffer size is quite large, $\mathtt{height}\cdot\mathtt{W\_SIZE}$,
as each thread needs its own memory buffer for each row.
The exact mapping of thread lanes and row numbers into \texttt{buf} is arbitrary,
as long as it is one-to-one,
and affects the number of shared memory bank conflicts and the efficiency of the parallel reduction step.
The mapping presented in Algorithm~\ref{alg:CMRS}, i.e. a cyclic assignment of threads, is designed to minimize the latter factor.
For example, for $\mathtt{height} = 8$ one can reduce $32\times8$ partial row sums in \texttt{buf}
into 8 row sums using just 9 instructions.

Our implementation needs \texttt{height}-fold more shared memory than the vector kernel of Ref.\
\cite{Bell09}.
On the one hand this is beneficial for the parallel reduction, but on the other hand it imposes a
severe limit on acceptable values of \texttt{height}, as the buffer size in currently available  GPUs is restricted to 48~KB.
For example, if we take $\mathtt{height} = 16$ and store the data as 4-byte numbers, 64 bytes of the shared
memory will be needed for each thread, and so the maximum number of resident threads per multiprocessor
will be limited to 768, which translates into the occupancy of 50\% for the Fermi and only 37.5\% for the Kepler architecture.
Taking into account  that a large number of resident threads is necessary to hide large memory latencies,
we can safely assume that  the maximum value of $\mathtt{height}$ in an  efficient implementation
for Fermi- or Kepler-class GPUs does not exceed 16.
This, in turn, implies that the values stored in array \texttt{RowInStrip} are in the range
0,\ldots,15 and hence can be encoded in just \texttt{CMRS\_BITS} = 4 bits. The remaining
28 bits are enough to store column indices of matrices with less than $2^{28}$ columns.
This is $\approx$ 20 times more than the size of the largest sparse matrix that we were able to test
on a 6~GB device.
The SpMV kernel on GPUs is so much memory-bound that it is of utmost importance
to reduce its memory footprint, even at the cost of several arithmetic operations, which in this kernel are
almost free, hence the idea of compressing two integers into a single 32-bit word.

We also implemented several optional performance  optimizations.
The first one consists in buffering the input vector in the
texture cache \cite{Bell09} or the new 48K read-only cache \cite{CUDA55} rather than in the L1 cache.
The second one consists in enlarging the shared memory size
from 16 to 48~KB, at the cost of the L1 cache size.
The third optimization strategy, adapted from \cite{Baskaran2009},
aims at improving the effective memory bandwidth for arrays \texttt{Val} and
\texttt{ColInd}, for large $\mu = \mathtt{nnz}/\mathtt{rows}$, by first accessing the non-aligned portion of a strip
and then accessing the remaining, aligned portion at full speed.
The fourth one consists in reordering the elements of CMRS arrays so that  the
index array \texttt{ColInd} is first sorted by strip indices and then within the same strip by column
indices. The idea behind such ordering is the same as for the row-major ordering in the CRS format: enhance the
frequency of coalesced or cache-buffered accesses of a SIMD unit to the elements of the input vector.
Note that most matrices coming from real problems have some internal structure and the locations of nonzero elements in
neighboring rows are correlated. In such cases reordering the entries in the CMRS arrays can have a pronounced impact
on SpMV efficiency.
Assuming again $\mathtt{height} = 2$, the CMRS representation of $\hat M$ after data reordering would read:
\begin{eqnarray*}
\texttt{Val}    &=& \left[\; 1 \;\; 3 \;\; 2 \;\; 4 \;\; 5 \;\; 7 \;\; 8 \;\; 6 \;\; 9 \;\; 10 \;\right], \\
\texttt{ColInd} &=& \left[\; 0 \;\; 1 \;\; 3 \;\; 4 \;\; 2 \;\; 2 \;\; 3 \;\; 4 \;\; 4 \;\; 4 \;\right], \\
\texttt{RowInStrip} &=& \left[\; 0 \;\; 1 \;\; 0 \;\; 1 \;\; 0 \;\; 1 \;\; 1 \;\; 0 \;\; 1 \;\; 0 \;\right], \\
\texttt{StripPtr} &=& \left[\; 0 \;\; 4 \;\; 9 \;\; 10 \;\right],
\end{eqnarray*}
Note that the reordering affects only the matrix internal representation and
does not involve any actions on the input and output vectors. Moreover, reordering is
local to strips and hence is prone to parallelization.

%%%%%%%%%%%%%%%%%%%%%%%%%%%%%%%%%%%%%%%%%%%%%%%%%%%%%%%%%%%%
%%%%%%%%%%%%%%%%%%%%%%%%%%%%%%%%%%%%%%%%%%%%%%%%%%%%%%%%%%%%
%%%%%%%%%%%%%%%%%%%%%%% SECTION %%%%%%%%%%%%%%%%%%%%%%%%%%%%
%%%%%%%%%%%%%%%%%%%%%%%%%%%%%%%%%%%%%%%%%%%%%%%%%%%%%%%%%%%%
%%%%%%%%%%%%%%%%%%%%%%%%%%%%%%%%%%%%%%%%%%%%%%%%%%%%%%%%%%%%

\section{Performance Model}

\begin{figure}
\begin{center}
  \includegraphics[width=0.5\columnwidth]{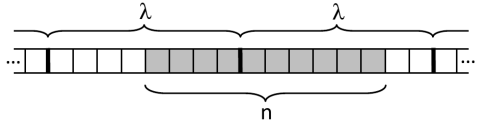}
  \caption{A warp transfers a chunk of $n$ consecutive bytes
  to or from the device accessing the memory in $\lambda$-byte-long segments.
  This can lead to the bandwidth waste and low kernel efficiency.
  }\label{fig:model_explained}
\end{center}
\end{figure}

The performance model of the CMRS format is based on a few simplifying assumptions:
the kernel is memory-bound;
each warp processes exactly $z$ nonzero matrix elements;
the data is read from or written to contiguous chunks of memory of size $n = bz$, where $b=4$ or $8$
is the number of bytes occupied by a data item;
finally, the number of memory transactions is equal to the number of distinct
memory segments of size $\lambda$ touched by the warp while accessing the
$n$-byte chunk of memory,
 see Fig.~\ref{fig:model_explained}.
If one also assumes that the beginning of the $n$-byte long chunk
is uncorrelated with the the
global memory segment boundaries,
 one concludes (see the Additional Material) that the ratio of the mean number bytes transferred
to the bytes actually requested by the kernel is
\begin{equation}
  \label{eq:def-f}
  f(h,\mu) = 1 + \frac{\lambda-b}{h \mu b},
\end{equation}
where $h$ denotes the strip height (note that $\mu$ = $z/h$ and $b/\lambda \ll 1$).
This number must be as close to 1 as possible for the kernel to be efficient.
For $h=1$ this formula estimates the efficiency of the vector kernel in the plain CRS format.
Thus, substituting  $ h=1, \lambda = 128$ \cite{CUDA55}, $ b=4$ (single precision), and $\mu = 2$
(only two nonzero matrix elements per row on average), one obtains $f = 16.5$, which means that for every 16 bytes
transferred by the CRS vector kernel, $\approx$ 15 are wasted. However, the CMRS kernel with $h=16$
would reduce $f$ down to $\approx 2$. Equation (\ref{eq:def-f}) can be used
to estimate the acceleration of the CMRS kernel over the plain CRS vector kernel,
\begin{equation}
  \label{eq:def-a}
  a(h,\mu) = \frac{f(1,\mu)}{f(h,\mu)} \approx 1 + \frac{h-1}{1 + h\mu b /\lambda}.
\end{equation}

Assuming that $b/\lambda = 1/32$, this formula suggests that for
extremely sparse matrices ($\mu \lesssim 5$) and small values of $h$
the CMRS format should be able to accelerate the plain CRS-based vector  kernel \texttt{height}-fold,
as in this case $a(h, \mu) \approx h$.
However, the advantages of the CMRS format are not expected to be particularly high for $\mu\gtrsim 100$.
Moreover, for $h\gtrsim16$ the value of $\partial a/\partial h$ is rather small,
which gives a theorethical justification of setting 16 as the upper bound for \texttt{height} in our implementation.
This formula implies also
that processing several matrix rows with a single warp will
become even more critical if the value of $\lambda$ increases in some future GPU architectures.

%%%%%%%%%%%%%%%%%%%%%%%%%%%%%%%%%%%%%%%%%%%%%%%%%%%%%%%%%%%%
%%%%%%%%%%%%%%%%%%%%%%%%%%%%%%%%%%%%%%%%%%%%%%%%%%%%%%%%%%%%
%%%%%%%%%%%%%%%%%%%%%%% SECTION %%%%%%%%%%%%%%%%%%%%%%%%%%%%
%%%%%%%%%%%%%%%%%%%%%%%%%%%%%%%%%%%%%%%%%%%%%%%%%%%%%%%%%%%%
%%%%%%%%%%%%%%%%%%%%%%%%%%%%%%%%%%%%%%%%%%%%%%%%%%%%%%%%%%%%

\section{Results\label{sec:results}}

\subsection{Hardware and software specification}

The tests were performed on two Nvidia devices, GTX 480 (1.5~GB, ``Fermi'' architecture)
and Tesla K20M (5~GB. ``Kepler'' architecture).
The ECC memory support in the Tesla device was switched off for a larger bandwidth.
In both cases the operating system was a 64-bit Linux with Nvidia GPU driver v.~319.21 and CUDA 5.5.
Theoretical
capabilities of these devices are listed in Table \ref{tab:1}.

\begin{table}
\begin{center}
\caption{Theoretical peak capabilities of the devices used in tests  \label{tab:1} }
\begin{tabular}{lrrr}
\hline
  & GTX 480 & K20M\\
\hline
single prec. perform. [Tflop/s]    &   1.3   &     3.5 \\
double prec. perform. [Tflop/s]    &   0.17  &     1.2 \\
memory bandwidth [GB/s]            &  177    &    208 \\
\hline
\end{tabular}
\end{center}
\end{table}

%%%%%%%%%%%%%%%%%%%%%%%%%%%%%%%%%%%%%%%%%%%%%%%%%%%%%%%%%%%%
%%%%%%%%%%%%%%%%%%%%% SUB-SECTION %%%%%%%%%%%%%%%%%%%%%%%%%%
%%%%%%%%%%%%%%%%%%%%%%%%%%%%%%%%%%%%%%%%%%%%%%%%%%%%%%%%%%%%

\subsection{Test matrices}

The tests were performed using 132 square real matrices from the University of Florida Sparse Matrix Collection
satisfying $10^6 \le \mathtt{nnz} \le 10^8 $, including all matrices with $\mathtt{nnz} \ge 5\times10^6$,
 and three additional sparse matrices of our choice.
The matrices are of various sizes and represent a wide spectrum of applications and structure patterns.
In particular, our tests include all symmetric  matrices used in several recent studies on GPU SpMV performance
\cite{Garland:Spa2008,Bell09,Vazquez2009,Vazquez2010,Monakov10,Grewe2011}.
We excluded from the tests a few sparse matrices with dense rows
(\texttt{lp1}, \texttt{circuit5M}, \texttt{Chebyshev4}, \texttt{rajat30}, \texttt{FullChip}),
because such matrices require special algorithms, as will be discussed below.
Some UF SMC matrices, e.g. \texttt{shipsec8}, are available in two versions:
with and without explicit zero entries. In such cases we tested both representations if
the number of explicit zeros is larger than \texttt{nnz}/10.
The additional matrices of our choice include a synthetic matrix, \texttt{p7},  which is a large ($10^7\times10^7$)
random permutation matrix,  \texttt{dense4}, which is a dense $10^4\times10^4$ matrix treated as a sparse one,
and \texttt{aorta}, which is a sparse matrix representing the pressure equation
in the problem of the flow through the  human abdominal aorta
\cite{Malecha11}. We included \texttt{p7} to get a better insight into the role played by structural
correlations between adjacent rows and the impact of (un)coalesced accesses to the input vector;
\texttt{dense4} is an example of a matrix which can be processed at the highest performance;
and \texttt{aorta} is an example of a sparse matrix for which efficiency of the SpMV kernel is of utmost importance,
as it directly affects the time to solution in biomedical applications.

%%%%%%%%%%%%%%%%%%%%%%%%%%%%%%%%%%%%%%%%%%%%%%%%%%%%%%%%%%%%
%%%%%%%%%%%%%%%%%%%%% SUB-SECTION %%%%%%%%%%%%%%%%%%%%%%%%%%
%%%%%%%%%%%%%%%%%%%%%%%%%%%%%%%%%%%%%%%%%%%%%%%%%%%%%%%%%%%%

\subsection{Optimal CMRS parameters\label{sub:optimal-choice}}
For each test matrix and each combination  of optimization parameters, our CMRS implementation of the SpMV product was
called 11 times and the execution times were recorded.
The largest time was omitted
and the remaining 10 results were analyzed to find their average and standard deviation.
The optimization parameters included the strip height ($\mathtt{height}=2,3,\ldots,16$),
the number of threads per block ($\mathtt{BS} = 64j, j = 1,\ldots,8$), and
four Boolean parameters referring to the optimization techniques described in Sec.~\ref{sec:implementation},
independently for K20M (Kepler) and GTX~480 (Fermi) GPUs.
We searched this parameter space for universal values that would give SpMV times as close as possible
to the shortest SpMV execution time $\tau_\mathrm{min}$ obtained through the brute-force search,
for as many test matrices as possible. We came to the following conclusions. The optimal block size is 128 threads.
The optimal strip height depends on the GPU architecture and the number of bytes occupied by each matrix value and
reads 6 (K20M, float), 4 (K20M, double), 12 (GTX~480, float), or 8 (GTX~480, double). The data should be sorted, the
size of the shared memory per multiprocessor should be set to the maximum value (48~KB), the input vector should be cached
either in the texture cache (GTX~480) or, if supported by the device, in the new 48~KB read-only cache (K20M), and
arrays \texttt{Val} and \texttt{ColInd} should be aligned for reading if $\mathtt{nnz}/\mathtt{rows} \ge 32$.

One should bear in mind that all these optimization parameters are not only correlated with each other,
but also depend on the representation of the matrix values (float or double) and on the matrix structure. For example,
for some matrices the texture cache turns out more efficient than the much larger 48~KB cache
and quite often there exist better values of the strip height.
Fortunately, the above-mentioned choice of the optimization parameters yields
optimal or nearly optimal SpMV times for most of the matrices
(see Sec.~\ref{sub:tuning} below).

\begin{figure}
\centering
\includegraphics[width=0.475\textwidth]{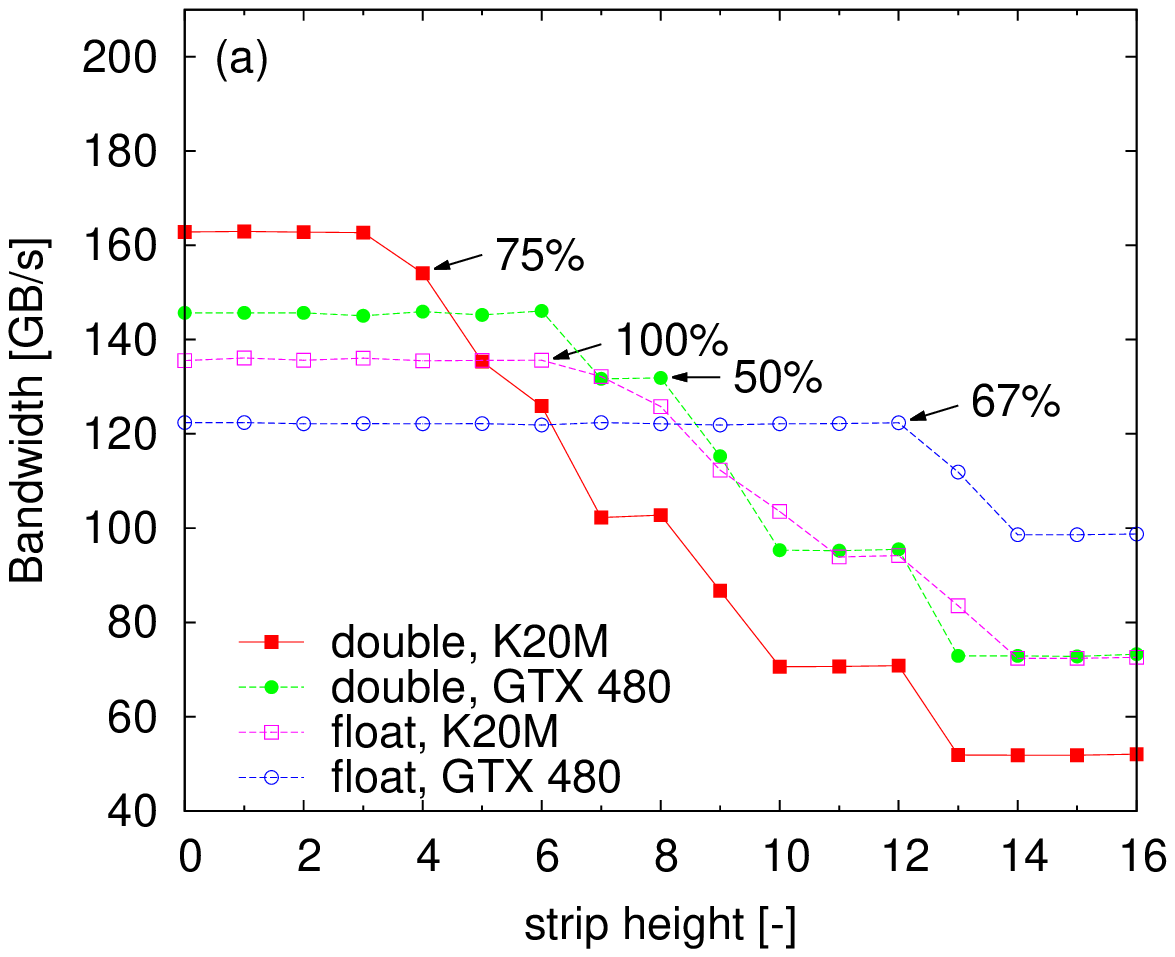} \hspace{1ex}
\includegraphics[width=0.475\textwidth]{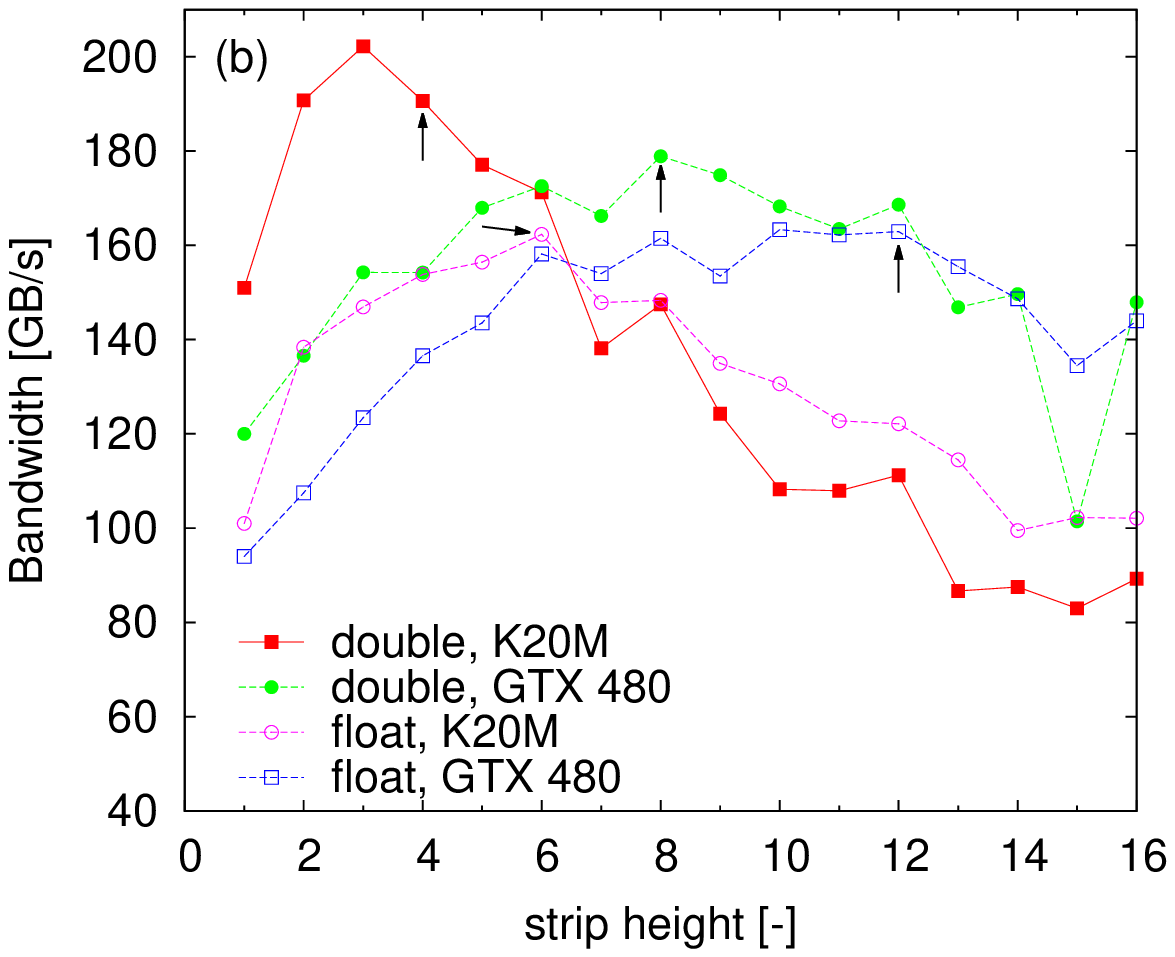}
 \caption{(Color online) The dependency of the memory bandwidth on the strip height for (a)~an idealized case of
  a kernel that only reads arrays \texttt{ColInd} and \texttt{Val} and
  (b) full CMRS SpMV kernel for matrix \texttt{rma10}
  calculated using the $\beta_-$ metric of Eq.~(\protect\ref{eq:beta}).
  Arrows point at the data obtained for the values of the optimal strip height,
  as adopted in our CMRS implementation. The arrow labels in panel (a)
  show the corresponding multiprocessor occupancy, whereas  ``float''
  and  ``double' refer to the type of elements stored in \texttt{Val}.
 \label{fig:rma10}}
\end{figure}

While the choice of the Boolean optimization parameters can be rather easily justified based on general
properties of GPUs, the values of the optimal block size and  strip height deserve closer inspection.
The value of \texttt{BS} = 128 is the smallest block size which allows for the full utilization of GPU's
memory bandwidth (data not shown), and small blocks are preferable for problems where different warps may have
to process different amounts of data.
As for the optimum value of the strip height, its value limits the occupancy, which, in turn, has
a profound impact on the kernel bandwidth. This is illustrated in Fig.~\ref{fig:rma10}.
Panel (a) shows how the speed of reading arrays \texttt{ColInd}
and \texttt{Val} depends on \texttt{height}.
We measured this relation using a very simple kernel that does nothing but read
simultaneously two streams of data from the arrays,
given that \texttt{height} words (floats or doubles) per thread
are reserved in the shared memory and limit the occupancy.
The results show some regression of the Kepler architecture relative to its predecessor. First, since
the maximum number of concurrent threads per multiprocessor was increased in Kepler by 4/3
without increasing the size of the shared memory, the number of shared memory words per thread
available in Kepler for a given target occupancy was lowered by 3/4.
Second, while lowering the occupancy down to 2/3 does not affect the speed at which Fermi can
access the global memory,
this speed deteriorates quickly in Kepler once the occupancy drops below 100\%. Consequently,
the optimal value of \texttt{height} for Kepler is expected  to be $4/3\times 3/2 = 2$ times lower than for Fermi,
which we can actually see in our tests.

The impact of the bandwidth-occupancy relation on the
actual performance of the CMRS SpMV kernel is visualised in Fig.~\ref{fig:rma10}~(b). Note that
as the value of \texttt{height} is increased, the kernel bandwidth initially quickly increases and either saturates
at \texttt{height} $\approx 8$ or hits the threshold value above which the GPU memory bandwidth starts to deteriorate.
From this point on the performance of the SpMV kernel starts to decrease,
as it is limited by the occupancy-related factors.
It is instructive to see how closely the curves in panel (b) follow those in panel (a),
especially for K20M.

%%%%%%%%%%%%%%%%%%%%%%%%%%%%%%%%%%%%%%%%%%%%%%%%%%%%%%%%%%%%
%%%%%%%%%%%%%%%%%%%%% SUB-SECTION %%%%%%%%%%%%%%%%%%%%%%%%%%
%%%%%%%%%%%%%%%%%%%%%%%%%%%%%%%%%%%%%%%%%%%%%%%%%%%%%%%%%%%%

\subsection{Methodology}
All computations were also repeated using two standard CRS-based SpMV kernels: scalar and vector,
as described in Sec.~\ref{sub:existing-implementations}. We used our own implementations of these kernels and
applied the brute-force method to find the best possible SpMV times.
The vector kernel was optimized with respect to all relevant parameters used for CMRS optimization, whereas
the scalar kernel was optimized with respect to the value of the block size and the
usage of the cache(s).
The purpose of using extensive brute-force search for the CRS data format
was to ensure that any acceleration of the CMRS over CRS implementation
is related to the data format.
Finally, we also measured the computational efficiency of three freely available SpMV implementations
for GPUs: Nvidia cuSparse 5.5 implementations for CRS and HYB formats
and the CUSP~0.3.0 implementation (CSR-tex) for the CRS format.
Each library function was treated as a black box and called using the default configuration.
CuSparse is a closed-source, proprietary library that can be regarded as an industry standard and reference point,
whereas CUSP is an open-source library containing several SpMV implementations for various data formats.
While we found that the CUSP kernels are generally less efficient than other kernels considered in this study,
we  decided to include the data for the CUSP CSR-tex kernel, as it features an improved version of the CRS-vector kernel,
aimed at accelerating the SpMV operation for extremely sparse matrices.

For each matrix the computational efficiency of the CMRS SpMV kernel was determined as the ratio of the  number
of elementary arithmetic operations to the
SpMV kernel time, i.e.\ $(2\mathtt{nnz}-\mathtt{rows})/\tau$. The bandwidth was
calculated as the total number of bytes that had to be transferred to or from the GPU main memory, $\beta$, divided by $\tau$.
We considered two extreme cases: the input vector either is not cached or is fully cached.
The bytes transferred in each case,  $\beta_-$ and $\beta_+$, respectively, are calculated from
 \begin{eqnarray}
  \label{eq:beta}
   \beta_- &=& (2s_\mathrm{v} + s_\mathrm{i})*\mathtt{nnz} + s_\mathrm{i} *\mathtt{strips} + s_\mathrm{v} * \mathtt{rows}, \nonumber \\
   \beta_+ &=& (s_\mathrm{v} + s_\mathrm{i})*\mathtt{nnz} + s_\mathrm{i} *\mathtt{strips} + 2s_\mathrm{v}*\mathtt{rows},
 \end{eqnarray}
where $s_\mathrm{v}$ is the size (in bytes) of data entries in \texttt{val} and $s_\mathrm{i} = \mathtt{sizeof(int)} = 4$.
The expression for $\beta_+$ is the number of bytes necessary to store the matrix and the input and output vectors.
The value of $\beta_-$ exceeds $\beta_+$ by $s_\mathrm{v}(\mathtt{nnz} - \mathtt{rows})$,
for if the cache is absent, reading elements of the input vector requires $\mathtt{nnz}$ rather than
$\mathtt{rows}$ transfers of the input vector components.

The bandwidth can be defined either as $\beta_-/\tau$ or $\beta_+/\tau$,
which leads to two definitions of memory utilization efficiency, $\eta_\pm$:
\begin{equation}
 \label{eq:def:eta}
   \eta_\pm = \frac{\beta_\pm}{\tau} \frac{1}{B},
\end{equation}
where $B$ is the theoretical hardware bandwidth of the device. Clearly,
$\eta_+ < 1$ and a value of $\eta_- \ge 1$ indicates
that the device efficiently buffers the input vector in its caches.

%%%%%%%%%%%%%%%%%%%%%%%%%%%%%%%%%%%%%%%%%%%%%%%%%%%%%%%%%%%%
%%%%%%%%%%%%%%%%%%%%% SUB-SECTION %%%%%%%%%%%%%%%%%%%%%%%%%%
%%%%%%%%%%%%%%%%%%%%%%%%%%%%%%%%%%%%%%%%%%%%%%%%%%%%%%%%%%%%

\subsection{SpMV multiplication results}
\begin{figure}
\centering
\includegraphics[width=0.9\textwidth]{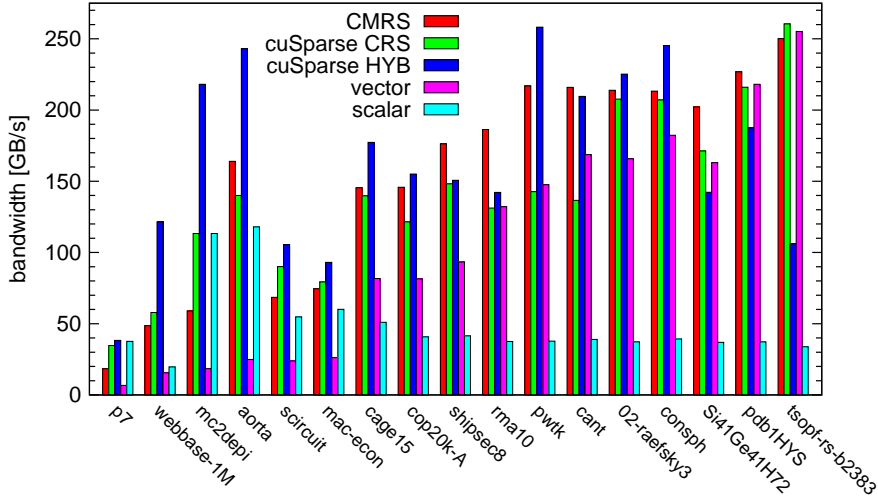}
 \caption{(Color online) Effective memory bandwidth $\beta_-/\tau$ of five SpMV kernels on the Kepler K20M GPU for selected
  sparse matrices in double precision representation. The matrices are ordered according to $\mu$.
 \label{fig:2}}
\end{figure}

The memory bandwidth $\beta_-/\tau$ for five SpMV kernels running on the Kepler K20M GPU
for selected matrices (double precision) is presented in Fig.~\ref{fig:2}.
The matrices in this figure include all square matrices used in Ref.~\cite{Bell09}.
They were ordered according to the average row length, $\mu$,
which ranges from 1 (matrix \texttt{p7}) to $\approx 484$ (\texttt{tsopf-rs-b2383}).
In this set, matrix \texttt{p7} constitutes an extreme case in which accesses to the input vector are totally uncoalesced.
Moreover, since in this case the CMRS algorithm uses the value of \texttt{height} = 4,
only 4 of the 32 threads in a warp are actively processing the matrix elements.
This leads to very inefficient memory bandwidth utilization, with $\eta_+=\eta_- \approx 0.09$.
Matrix \texttt{tsopf-rs-b2383} constitutes another extreme case in which
accesses to the input vector are well coalesced. The memory bandwidth attained
for this matrix by the CMRS kernel
is 250~GB/s, which yields $\eta_-= 1.22$ and $\eta_+= 0.73$.  The fact that $\eta_- > 1$ indicates that
K20M can efficiently buffer the input vector in its caches. However, the efficiency of the previous generation GPU, GTX~480,
turned out to be  even better for this matrix
($\eta_-= 1.40$, $\eta_+= 0.80$, 248~GB/s), even though GTX~480 has a smaller L2 cache,
no 48~KB read-only cache, and is 7 times slower at double precision arithmetics (c.f.\ Tab.~\ref{tab:1}).
Note that pre-Fermi GPUs, e.g.\ GTX~285,
which had neither L1 nor L2 caches, allowed for far less efficient data caching ($\eta_- \le 1.08$) \cite{Bell09}.

\begin{figure}
\centering
 \includegraphics[width=0.49\textwidth]{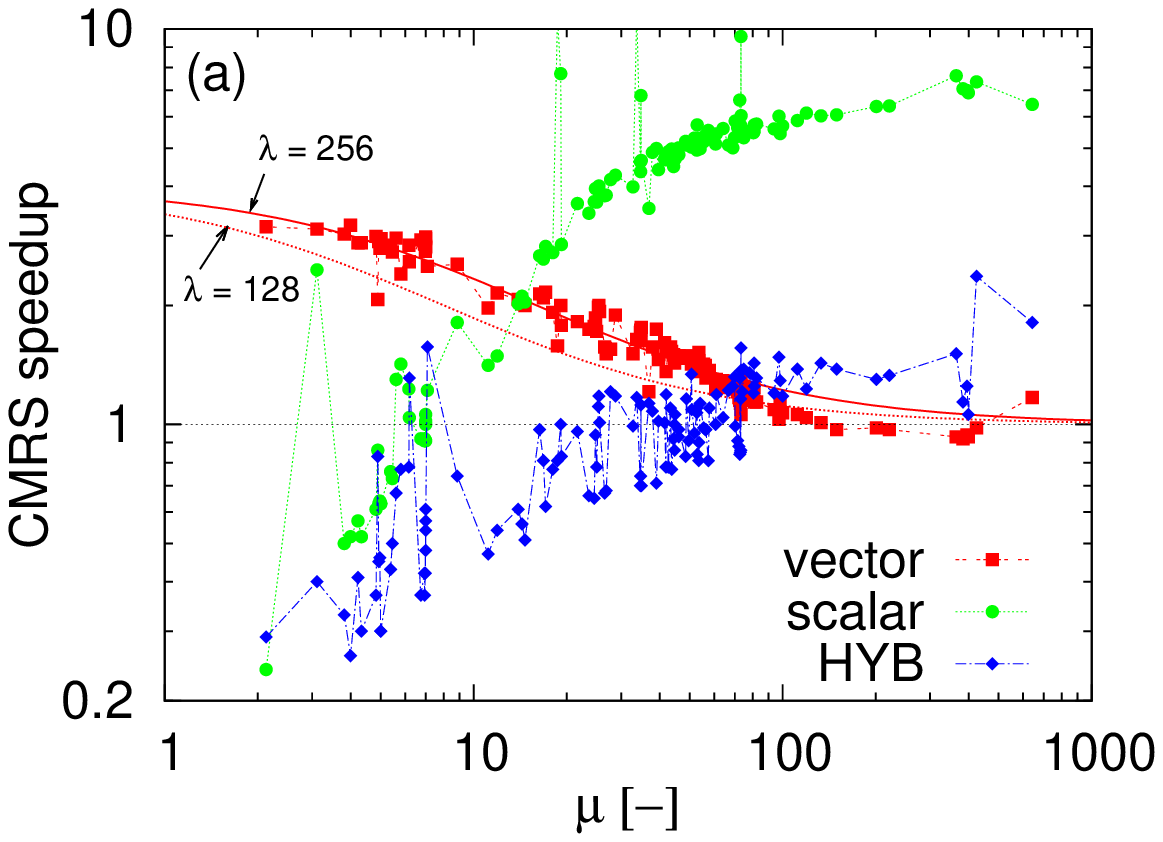}
 \includegraphics[width=0.49\textwidth]{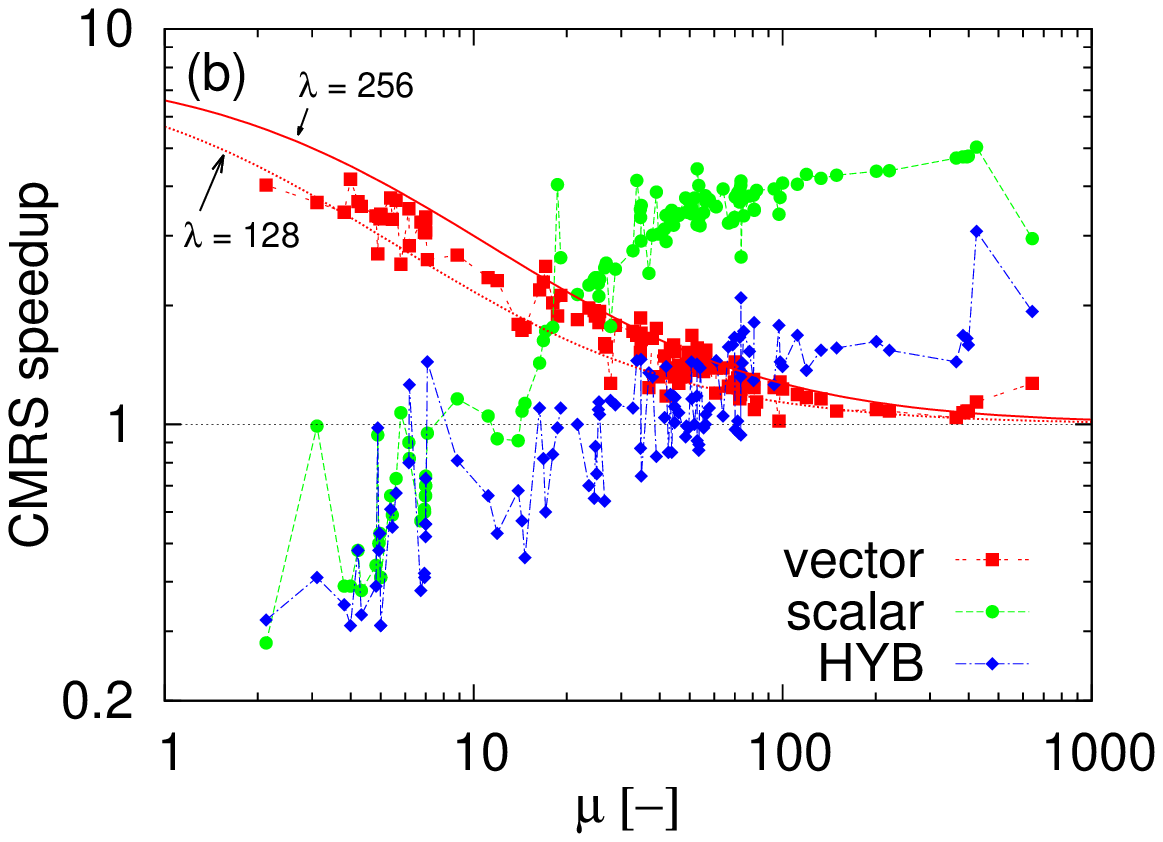}
  \caption{(Color online)
   Performance of the CMRS SpMV kernel relative to three other kernels for
   the test matrices as a function of the mean row length, $\mu$,
   for (a)~K20M  and (b) GTX~480 in double precision.
   The smooth lines were computed from
   Eq.~(\ref{eq:def-a}) with  $\lambda=128,256$.
   \label{fig:3}}
\end{figure}

To compare different SpMV algorithms, we  analysed the results obtained for the
sparse matrices from UF SMC, assuming that this collection contains a representative sample of sparse matrices.
The speedup of our implementation over three other SpMV kernels, vector, scalar and hybrid,
for K20M (double precision) and GTX~480 (single precision) is shown in Fig.~\ref{fig:3}.
Results for K20M in double precision ($\mathtt{height} = 4$) are interesting from the practical point of view,
whereas the results for GTX~480 in single precision ($\mathtt{height} = 12$) allow to estimate to what extent
the performance of the CMRS kernel is affected by the kernel occupancy.
Clearly, $\mu$ turns out to be a relevant parameter for determining relative performance of various SpMV implementations.
However, perhaps an even more striking feature of the two graphs is their similarity, which reflects the fact
that the performance of SpMV kernels is highly influenced by the matrix structure.
The scalar and hybrid kernels give the shortest SpMV times
for small $\mu$, but their efficiency decreases as $\mu$ is increased,
with the scalar kernel being very inefficient for large $\mu$.
This is related to the inability  of the scalar kernel to coalesce data transfers if $\mu$ is large.
The vector kernel behaves in just the opposite way: its efficiency relative to other kernels is very good for large $\mu$,
but it decreases as $\mu$ drops below $\approx 100$, as predicted by Eq.\ (\ref{eq:def-a}).

The smooth lines in Fig.~\ref{fig:3} show the speedup of the CMRS kernel over the vector kernel, as predicted by
Eq.~(\ref{eq:def-a}), for the data in double precision ($b=8$). We used two values of $\lambda$, 128
(as suggested by Nvidia for accessing contiguous streams of 4-byte data \cite{CUDA55}) and 256. For K20M
the agreement is very good for $\lambda=256$, whereas for the older architecture the experimental values appear to lie
between the two theoretical curves. The superiority of $\lambda=256$ for 8-byte data on the Kepler architecture
will be also discussed in Sec.~\ref{subsec:boost}.

\begin{figure}
\centering
\includegraphics[width=0.6\textwidth]{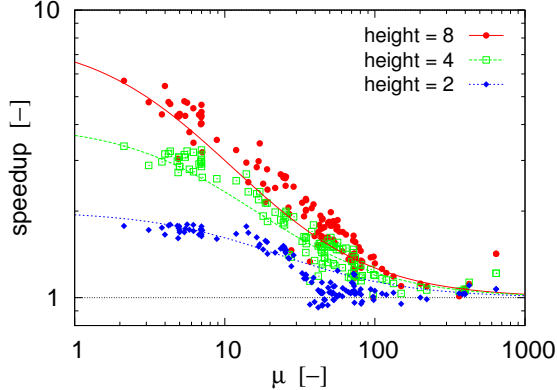}
\caption{(Color online)
 The speedup of the CMRS SpMV kernel over the vector kernel as a function of the mean row length,
 $\mu$, for GTX~480 in single precision and $\mathtt{height} = 2,4,8$
 (symbols) and
 the performance model predictions, Eq.~(\protect\ref{eq:def-a}) with $b=4, \lambda=128$ (lines).
\label{fig:4}}
\end{figure}

To better validate the performance model of the CMRS format, in Fig.~\ref{fig:4}
we compare its predictions for the 4-byte data with the results obtained for GTX~480 in single precision.
With this choice of the matrix value representation and the GPU architecture, the device runs at the full occupancy
 up to \texttt{height} = 8. As can be seen, the model
describes the actual speedup well.

\begin{figure}
\centering
\includegraphics[width=0.475\textwidth]{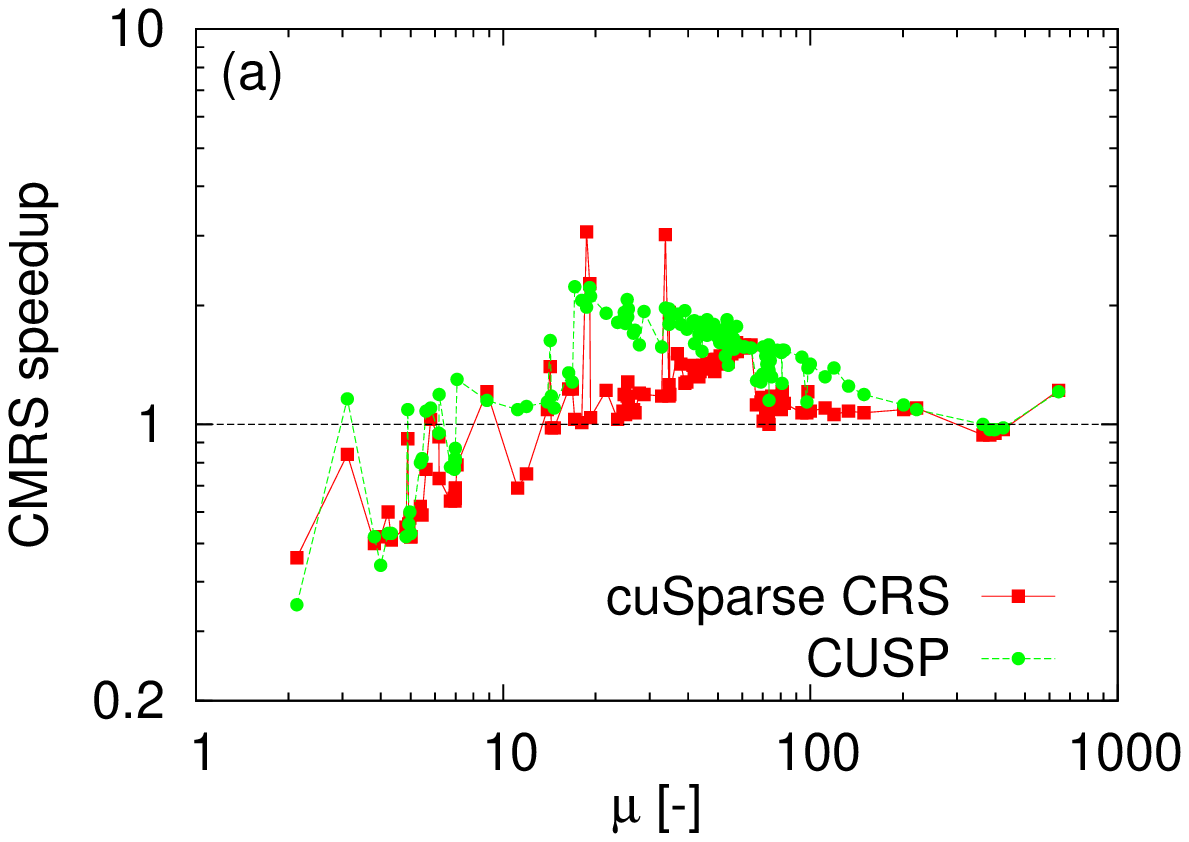} \hspace{1ex}
\includegraphics[width=0.475\textwidth]{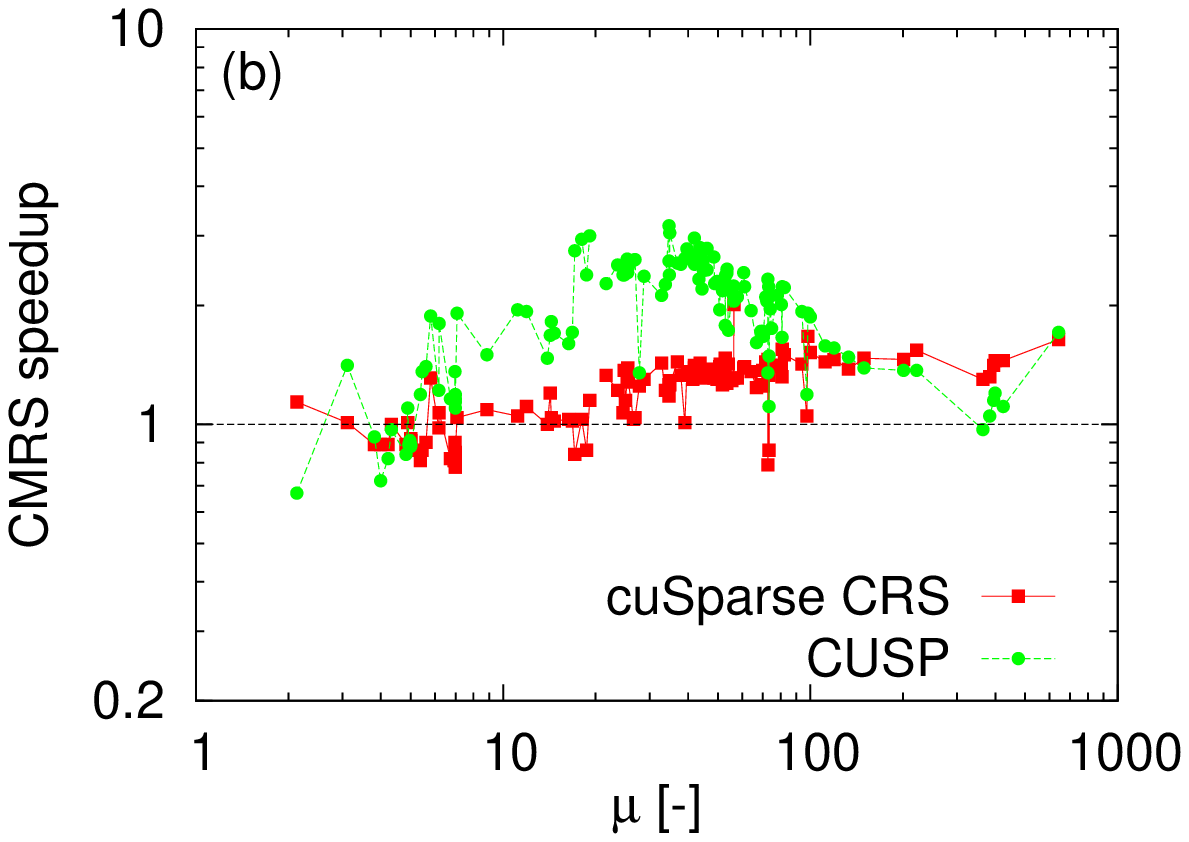}
 \caption{(Color online) The speedup of the CMRS SpMV kernel over the cuSparse CRS and CUSP CRS-tex kernels
  as a function of the mean row length,
  $\mu$, for (a)~K20M in double precision and (b) GTX~480 in single precision.
 \label{fig:cusp-susparse}}
\end{figure}

The speedup of our CMRS SpMV kernel over two remaining, CRS-based SpMV kernels,
cuSparse and CUSP, is shown in Fig.~\ref{fig:cusp-susparse}.
Again we show only
the data for the extreme cases of K20M in double precision (left panel)
and GTX~480 in single precision (right panel).
The cuSparse implementation turns out to be better optimized then CUSP and
our CMRS SpMV kernel  outperforms each of them for sufficiently large values of~$\mu$.

%%%%%%%%%%%%%%%%%%%%%%%%%%%%%%%%%%%%%%%%%%%%%%%%%%%%%%%%%%%%
%%%%%%%%%%%%%%%%%%%%% SUB-SECTION %%%%%%%%%%%%%%%%%%%%%%%%%%
%%%%%%%%%%%%%%%%%%%%%%%%%%%%%%%%%%%%%%%%%%%%%%%%%%%%%%%%%%%%

\subsection{Comparison of the computational efficiency of different SpMV kernels}
To compare the computational efficiency of different SpMV kernels,
we adopt a convention that implementation $A$ is significantly faster than $B$ if and only if
its execution time is at least 10\% shorter.

We found only one matrix (\texttt{kktpower}) for which the cuSparse 5.5 CRS
is significantly faster than any other SpMV kernel consider here.
Although for 26 matrices this kernel turns out significantly faster than our CMRS implementation, each of these matrices
is characterized by a low number of nonzero elements per row and for such matrices
the HYB kernel is usually even faster. A similar situation is observed for the scalar kernel, which
is significantly faster than any other SpMV kernel for only one matrix (\texttt{asia\_osm}).
The CUSP implementation is generally even
less efficient than \mbox{cuSparse}.
We found no matrix for which the vector kernel is significantly faster than any
other kernel.
Similar results were obtained for GTX~480 as well as for calculations in single precision.

The most interesting is comparison of our algorithm with the cuSparse 5.5 HYB implementation.
We found HYB to be significantly faster than any implementation (our implementation) for 55 (58) matrices.
On the other hand, our implementation is significantly faster than any other (HYB) implementation for 29 (46) matrices.
This is a good result, especially if one takes into account that
the HYB implementation
analyses the matrix structure and transforms it (e.g.\ by zero padding) accordingly
before the first SpMV routine can be called on it.
In Sec.~\ref{subsec:boost} we shall examine how techniques like zero-padding could be used to
further optimize the CMRS SpMV kernel.
Note also that the currently available implementation of the HYB format has rather high memory requirements.
For this reason the HYB implementation could not be run on GTX~480 for 14 largest matrices in double precision.

\begin{figure}
\centering
\includegraphics[width=0.475\textwidth]{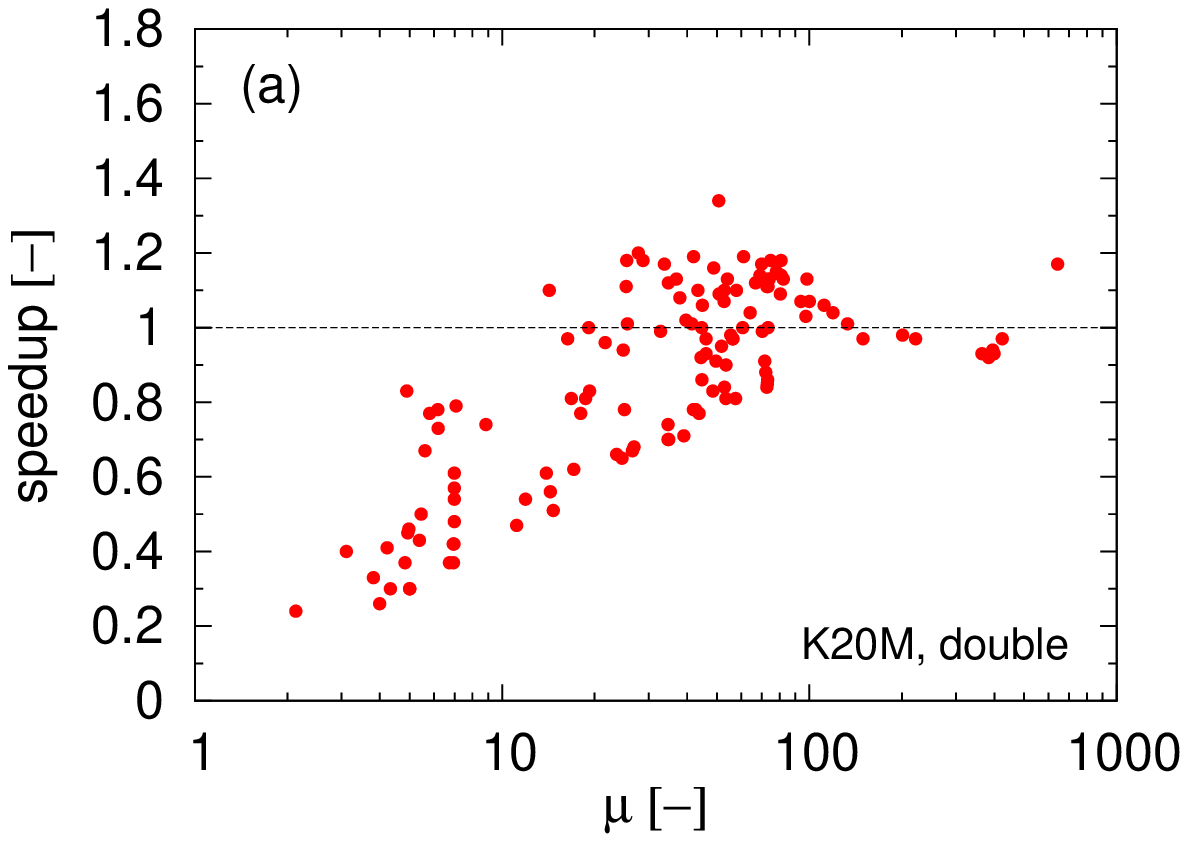} \hspace{1ex}
\includegraphics[width=0.475\textwidth]{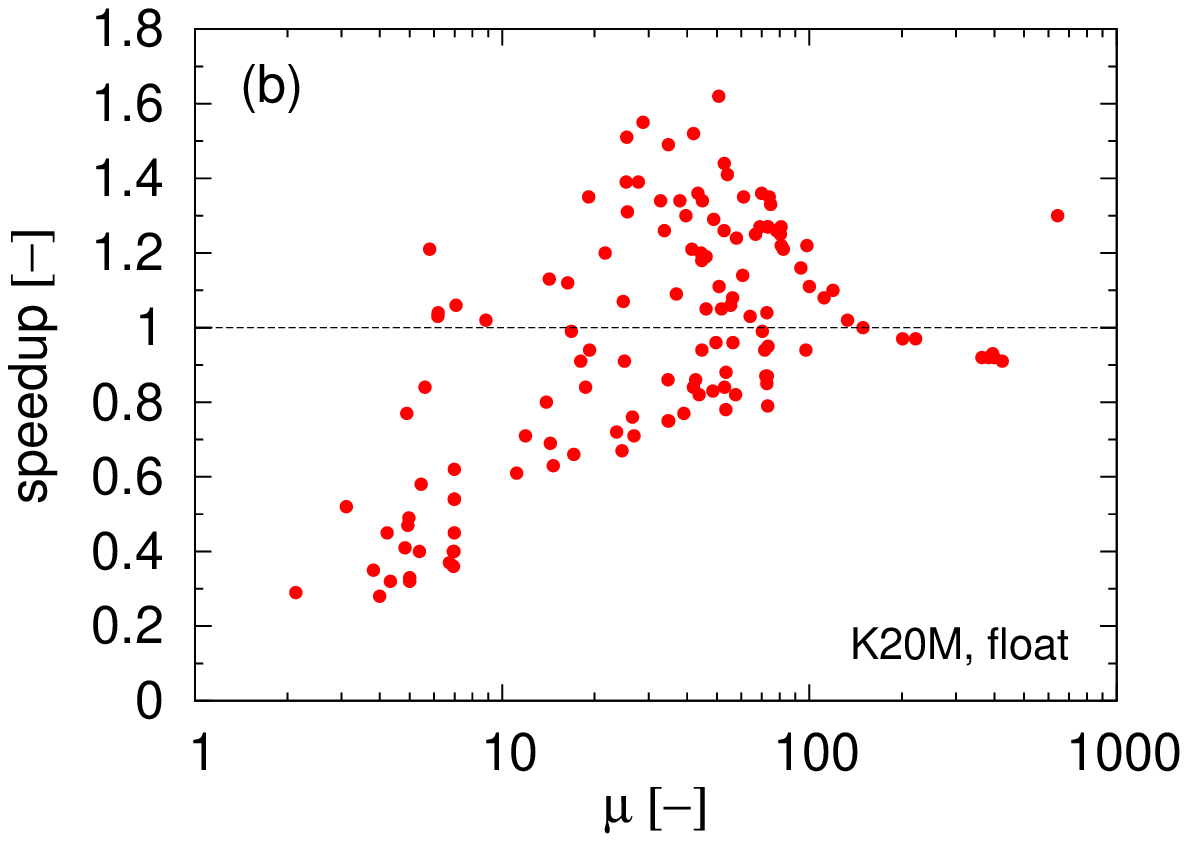}\newline
\includegraphics[width=0.475\textwidth]{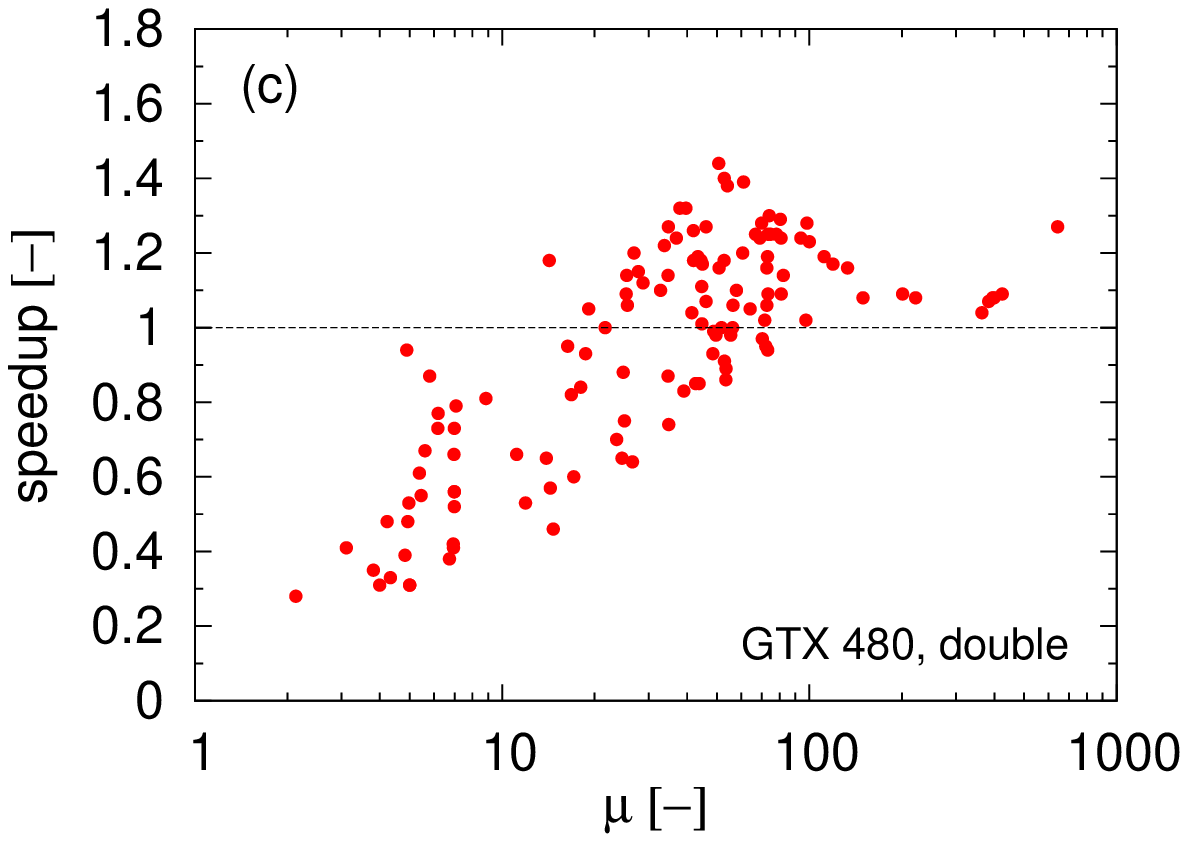} \hspace{1ex}
\includegraphics[width=0.475\textwidth]{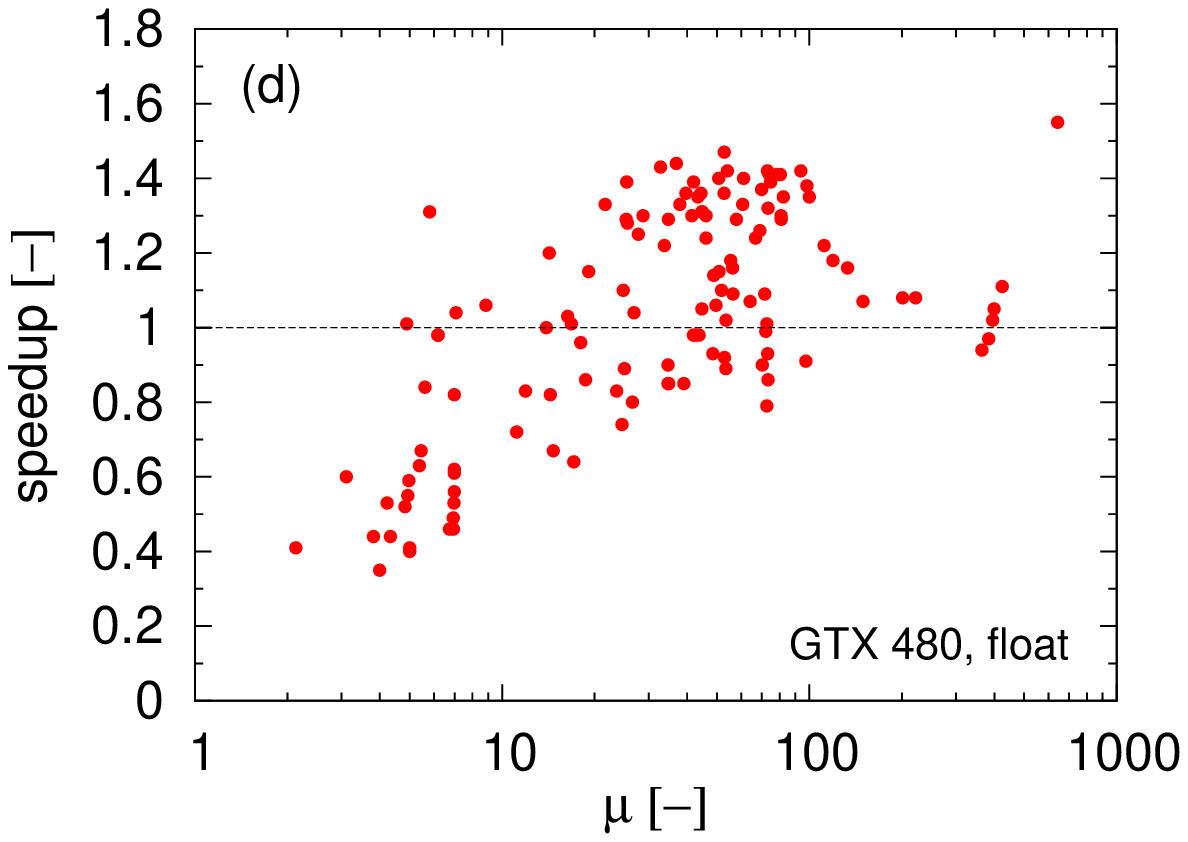}\newline
  \caption{(Color online) The speedup of our CMRS implementation over the best of all five
    alternative SpMV kernels as a function of the
    nonzero matrix elements per row ($\mu$), for K20M and GTX 480, in double and single precision.
   \label{fig:best}
  }
\end{figure}

From Fig.~\ref{fig:3} it can be immediately seen that
our CMRS implementation generally does not yield much
improvement over the vector implementation for $\mu \gtrsim 150$,
and tends to be systematically slower than HYB for $\mu \lesssim 20$.
Hence one expects that the advantages of CMRS will be most pronounced for moderate values of $\mu$.
This is confirmed by Fig.~\ref{fig:best},
which depicts
the speedup of our CMRS SpMV implementation
against the best of all five alternative SpMV implementations considered here, calculated individually for each matrix,
for the K20M and GTX~480 GPUs running in double and single precision mode. A striking similarity of the
results obtained for different architectures and different matrix value representations indicates that the
efficiency of an SpMV kernel depends mainly on the matrix structure. It is also clear that the efficiency
of our CMRS implementation in the most important case of the Kepler architecture (K20M)
in double precision is $\approx 10\%$ worse than for the Fermi architecture (GTX~480).
In particular, the largest speedup for K20M and GTX~480 is 34\%  and 44\%, respectively.
We believe this is an effect of the bandwidth-occupancy relation in GPUs, as discussed in Sec.~\ref{sub:optimal-choice}.
As might be expected, the CMRS format allows for even better
acceleration of the SpMV kernel if the calculations are performed in single precision.
The maximum speedup is $\approx 62\%$ for K20M and  $\approx 55\%$ for GTX~480,
even though in the former case we used a smaller value of the CMRS strip height.
We attribute this to the fact that the cuSparse 5.5 CSR kernel is apparently not well optimized for
the Kepler architecture in single precision (data not shown).

\begin{figure}
\centering
\includegraphics[width=0.475\textwidth]{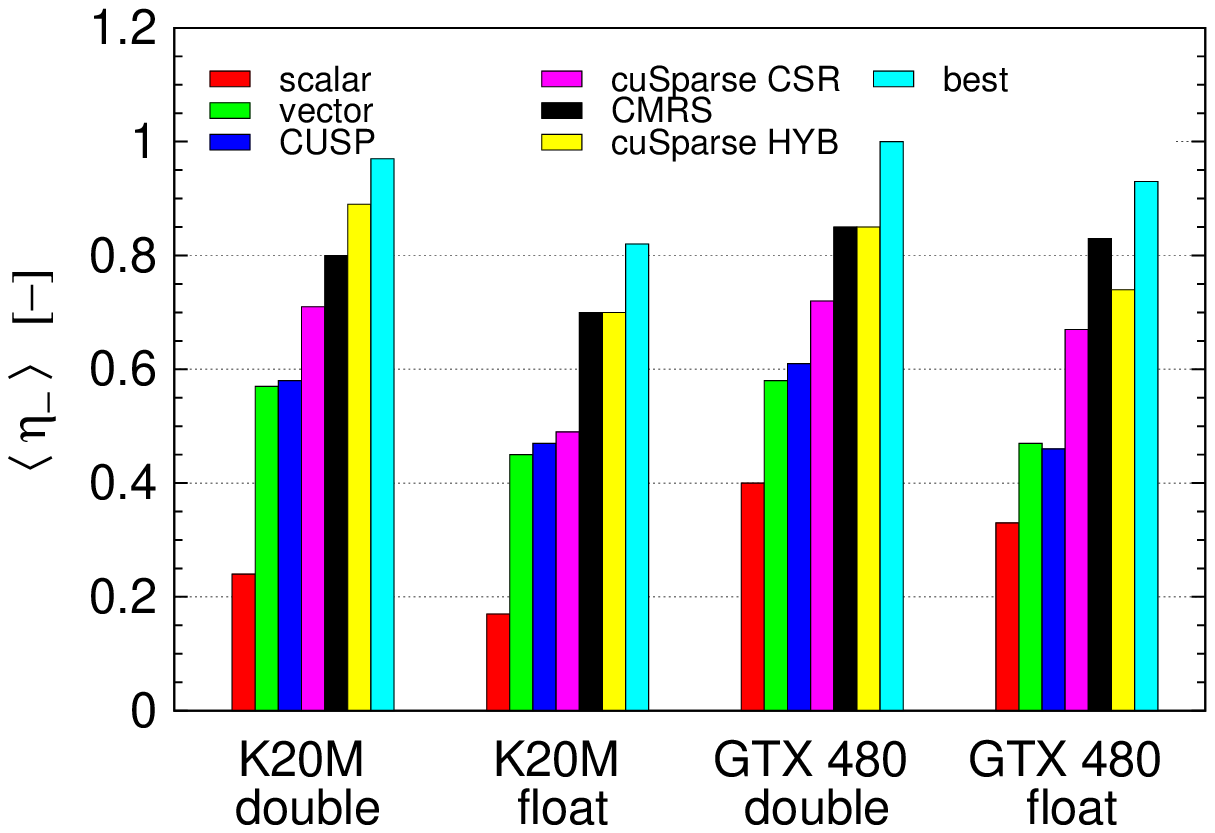} \hspace{1ex}
\includegraphics[width=0.475\textwidth]{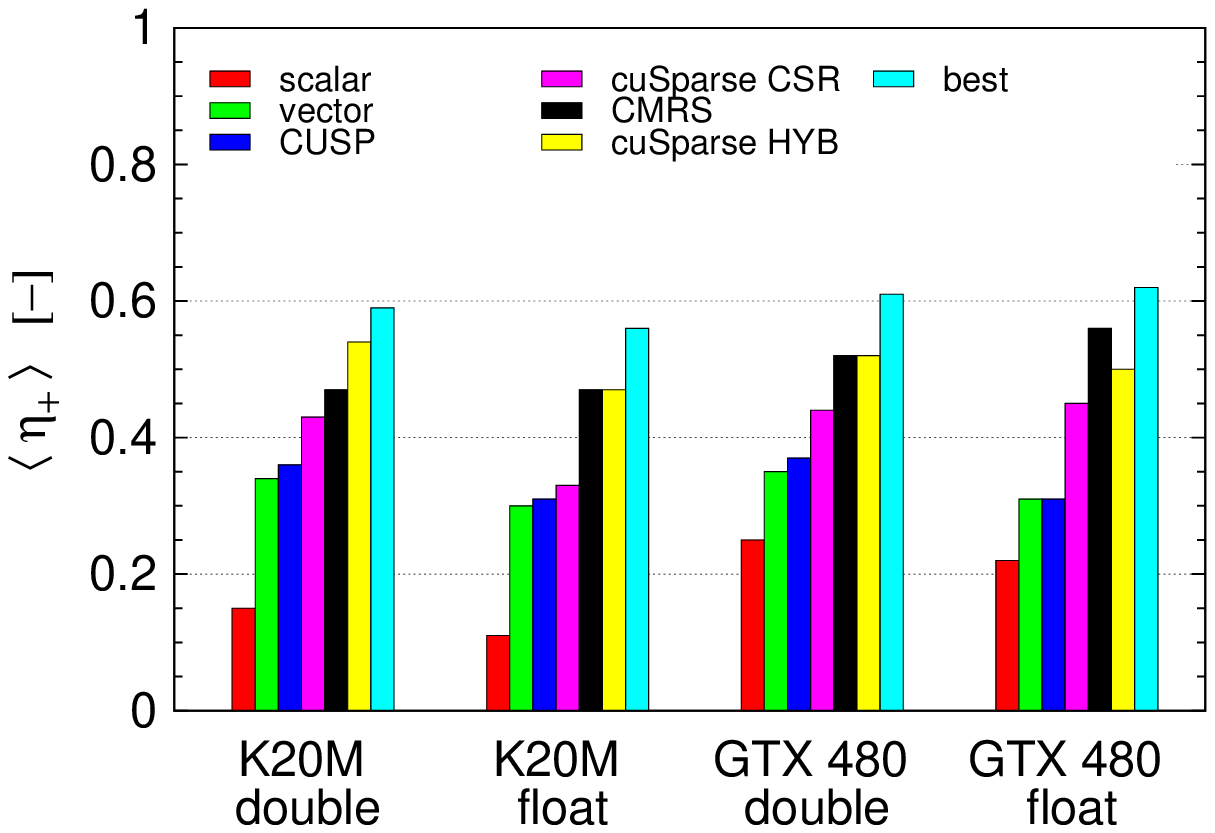}
 \caption{(Color online) The average memory utilization efficiency,
  $\langle \eta_- \rangle $ (left) and $\langle \eta_+\rangle$ (right), as defined in Eq.~(\ref{eq:def:eta}),
  for all tested SpMV kernels, GPU devices and matrix value representations,
  as a function of $\mu$. Label ``best'' denotes the results for the most efficient
  of all 6 SpMV kernels considered in this study, selected individually for each matrix.
 \label{fig:eta}}
\end{figure}

Since the SpMV operation is memory-bound,
efficiency of various implementations of this kernel can be compared using the memory utilization efficiency parameters
$\eta_\pm $, Eq.~(\ref{eq:def:eta}).
The results for all tested SpMV kernels, GPU devices and matrix value representations, averaged over
all tested matrices,  are shown in Fig.~\ref{fig:eta}.
We also included the results for a hypothetical kernel, denoted as ``best'',
in which the most optimal kernel is selected for a given sparse matrix
(in practice, this choice is limited to choosing between HYB and CMRS).
These results confirm that the scalar kernel is very inefficient as a general-purpose SpMV kernel, especially
in the newer (Kepler) architecture. Optimization of the cuSparse 5.5 CSR kernel appears to be unsatisfactory for
single precision arithmetics on K20M.  The best results, on average, are obtained for the cuSparse 5.5 HYB
and our CMRS implementation. Note that if one used the best kernel for a given matrix, $\langle\eta_-\rangle$
would rise to $\approx 1$ for double precision arithmetics on both Fermi and Kepler architectures,
which a very good result.

Since the value of $\eta_+$ is bounded from above by 1,
its value carries valuable information about the extent to which a kernel utilizes the hardware.
Its mean value for the best kernel is $\approx 0.6$ for both architectures,
which again should be considered as a very good result. Its value for individual matrices varies from
$\approx 0.13$ for permutation matrices to $\approx 0.80$ for the \texttt{af\_shell10} sparse matrix and
can be as high as $\approx 0.86$ for a dense $10~000\times10~000$ matrix treated as a sparse one.
A value of  $\eta_+$ much smaller than 1
might be used as an indicator that a significant performance boost could probably be
achieved through reordering of the matrix rows.

%%%%%%%%%%%%%%%%%%%%%%%%%%%%%%%%%%%%%%%%%%%%%%%%%%%%%%%%%%%%
%%%%%%%%%%%%%%%%%%%%% SUB-SECTION %%%%%%%%%%%%%%%%%%%%%%%%%%
%%%%%%%%%%%%%%%%%%%%%%%%%%%%%%%%%%%%%%%%%%%%%%%%%%%%%%%%%%%%

\subsection{Matrix transformation for better performance\label{subsec:boost}}
The basic CMRS format, as defined in Sec.~\ref{sec:CMRS},
allows for a quick and straightforward conversion to and from the CRS format without any
memory overhead. Can we relax these two conditions to allow for an even faster SpMV kernel?

A major issue with the implementation presented in Sec.~\ref{sec:implementation} is that it
requires each warp to reserve a \texttt{WARP\_SIZE}$\times$\texttt{HEIGHT} data array in the shared memory,
of which only \texttt{WARP\_SIZE} elements are utilized simultaneously.
In many  cases most of the shared memory may never be used by the warp that controls it.
The central problem is, however, that reducing the size of per-warp buffers in the shared memory
would allow to increase the value of \texttt{height}, which, following Eq.~(\ref{eq:def-a}),
should result in a significant kernel performance boost.

This problem can be coped with by changing the structure of the sparse matrix.
Here we briefly examine one such approach.
In Algorithm \ref{alg:CMRS} the buffer is accessed always through the same pattern: \texttt{buf[thread\_lane, r]}.
If we could replace it with \texttt{buf[thread\_lane \textrm{mod} M, r]},
where \texttt{M} $<$ \texttt{WARP\_SIZE},
the size of each per-warp buffer could be reduced to \texttt{M}$\times$\texttt{HEIGHT}, i.e.\ by a factor of
\texttt{WARP\_SIZE}/\texttt{M}. This will work provided that no threads
in a warp can access the same buffer location simultaneously. In most cases this condition
can be met by taking advantage of the fact that the CMRS format permits one to reorder the items in a strip
arbitrarily: it suffices to arrange the items in such a way that each warp
processes at most \texttt{M} items from a given row \texttt{r} and these items
are stored contiguously in the array. Such arrangement ensures that if the value of the row identifier \texttt{r} in
\texttt{buf[thread\_lane \textrm{mod} M, r]} is the same for some threads,
the values of the first indices into the buffer are different. Such arrangement can be, however, impossible
for some sparse matrices with highly variable row lengths, especially for small values of \texttt{M}.
In such cases the matrices must be filled with explicit
zeroes, which modifies the structure of the matrix internal representation.

We examined numerically the case \texttt{HEIGHT} = 16 and \texttt{M} = 8, which requires the same
buffer size as in the implementation analysed in the previous section for K20M and double precision,
but is characterized by a 4-fold larger value of the strip height. Since the value of \texttt{HEIGHT}
is now relatively high so that  each strip contains hundreds or even thousands of matrix elements,
we also applied another optimization: all strips were padded with zeroes to ensure the number of matrix items
they contain is a multiple of \texttt{WARP\_SIZE}. In this way all
memory accesses to arrays \texttt{Val} and \texttt{ColInd} are fully coalesced.
This comes at the cost of an additional modification of the internal matrix representation,
which in some cases may result in a noticeable memory overhead.
For 4 matrices,  the implementation considered here turned out to be significantly slower than that defined in
Sec.~\ref{sec:implementation} and we excluded them from further analysis.

\begin{figure}
\centering
\includegraphics[width=0.45\textwidth]{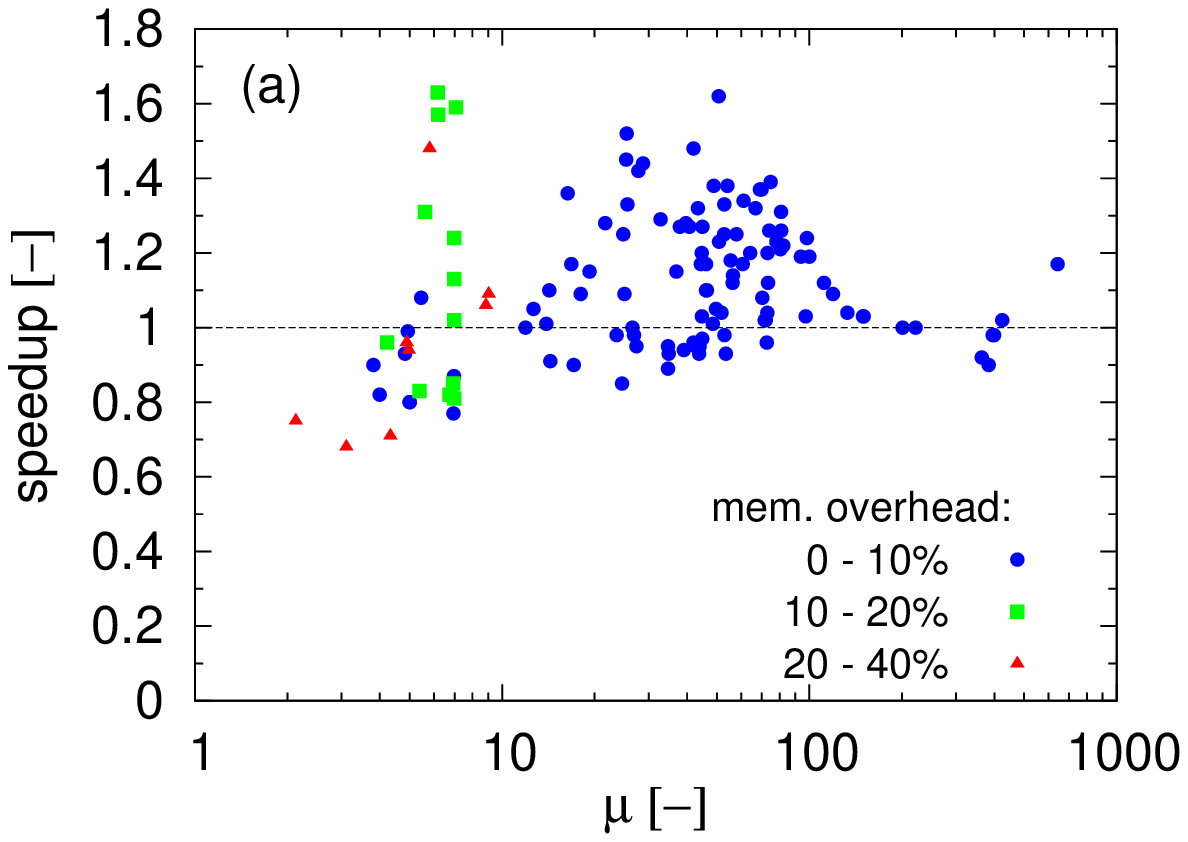}
\includegraphics[width=0.45\textwidth]{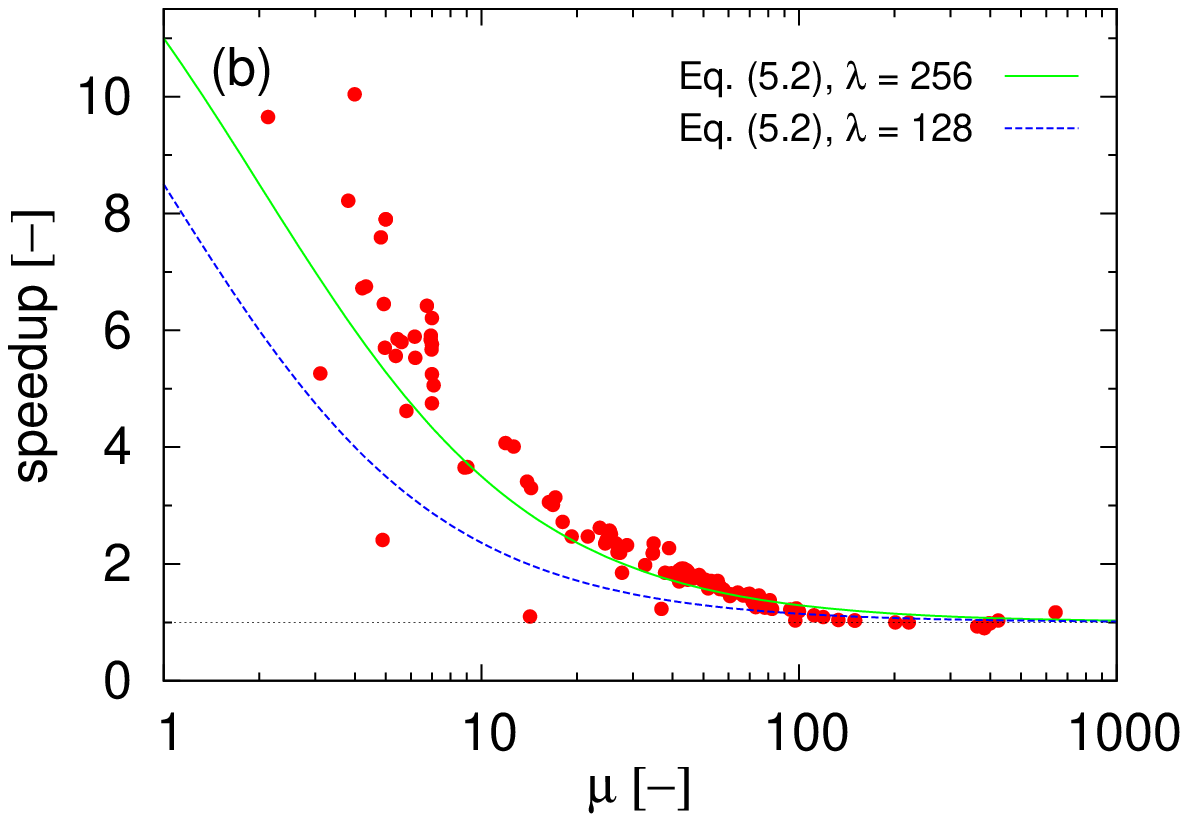}
 \caption{(Color online) The speedup of the modified CMRS SpMV kernel over (a) the best of all five
    alternative SpMV kernels and (b) the vector kernel as a function of the
     nonzero matrix elements per row ($\mu$), for K20M in double precision. Different symbols in panel (a) represent different
     levels of memory overhead related to padding the internal matrix representation with explicit zeroes.
 \label{fig:super-fast}}
\end{figure}

The results are shown in Fig.~\ref{fig:super-fast}a,
which corresponds directly to Fig.~\ref{fig:best}a.
Clearly, padding the CMRS matrix with explicit zeroes can result in a considerable
shortening of the SpMV execution time, especially for small values of $\mu$,
as expected from Eq.~(\ref{eq:def-a}). For example  the new algorithm turned out to
be 3 times faster than the implementation presented in Sec.~\ref{sec:implementation} for matrix \texttt{mc2depi}.
Our modified implementation is significantly
faster than any of five alternative SpMV kernels for 58 matrices,
with the greatest relative speed-up reaching 1.63 for the matrix \texttt{mac\_econ\_fwd500}, and
significantly slower than an alternative solution for only 13 cases, the worst case being \texttt{webbase-1M} for which
the execution time relative to the HYB kernel is 0.68.
The average memory efficiency is $\eta_- \approx 1.0$ and $\eta_+ \approx  0.6$, c.f.\ Fig.~\ref{fig:eta}.
Note that the zero padding introduced a significant memory overhead
only for some matrices with $\mu\lesssim10$.
Thus, the main advantage of  CMRS over HYB is that
it allows for an efficient single-kernel implementation without a significant memory overhead.
In contrast to this, HYB attempts to strike the balance between
a format that is computationally efficient but often requires a prohibitive memory overhead (ELL)
and a format that imposes no memory overhead, but is computationally inefficient (COO).

Figure~\ref{fig:best}b shows the speedup of the modified algorithm over the vector kernel for sparse matrices
in double precision on K20M and compares it
with the model, Eq.~(\ref{eq:def-a}), with two values of $\lambda=128,256$.
Clearly, the fit is much better for $\lambda=256$.

%%%%%%%%%%%%%%%%%%%%%%%%%%%%%%%%%%%%%%%%%%%%%%%%%%%%%%%%%%%%
%%%%%%%%%%%%%%%%%%%%% SUB-SECTION %%%%%%%%%%%%%%%%%%%%%%%%%%
%%%%%%%%%%%%%%%%%%%%%%%%%%%%%%%%%%%%%%%%%%%%%%%%%%%%%%%%%%%%

\subsection{Potential for further performance tuning\label{sub:tuning}}
Performance of many SpMV kernels can be significantly improved by adjusting
kernel optimization parameters to both the structure of the matrix and
the hardware on which the kernel is to be executed \cite{Choi:2010,Liu:2013,Vuduc:OSKI,Vuduc:Aut2003}.
For example,
Vuduc \cite{Vuduc:Aut2003} showed an up to four-fold acceleration
for modern cache-based superscalar machines.
However, finding optimal optimization parameters is usually costly and hence is often performed ``off-line''.
In particular, Choi et al.\ \cite{Choi:2010} developed efficient autotuning techniques for
their BELLPACK sparse matrix format and tested it on pre-Fermi GPUs.
However, unlike BELLPACK, CMRS does not use explicit storage of dense blocks to compress
the data structure and hence requires a different optimization strategy.

By comparing the default CMRS kernel times with those obtained for the same kernel launched
with the optimal parameters determined by the brute-force search,
we found that while the tuning of the default CMRS parameters is possible,
it is not expected to give a spectacular performance boost. For example,
the CMRS parameters could be tuned to speed up the kernel by at least 10\% for only
only 14 (16) test matrices on K20M (GTX 480) in double precision, and the maximum
acceleration was 18\% and 36\% for K20M and GTX~480, respectively.
Our preliminary results presented in the previous section indicate that
better results can be obtained by modifying the internal structure of the  sparse matrix,
e.g.\ by a suitable zero-padding.
Further research on this issue is necessary.

%%%%%%%%%%%%%%%%%%%%%%%%%%%%%%%%%%%%%%%%%%%%%%%%%%%%%%%%%%%%
%%%%%%%%%%%%%%%%%%%%%%%%%%%%%%%%%%%%%%%%%%%%%%%%%%%%%%%%%%%%
%%%%%%%%%%%%%%%%%%%%%%% SECTION %%%%%%%%%%%%%%%%%%%%%%%%%%%%
%%%%%%%%%%%%%%%%%%%%%%%%%%%%%%%%%%%%%%%%%%%%%%%%%%%%%%%%%%%%
%%%%%%%%%%%%%%%%%%%%%%%%%%%%%%%%%%%%%%%%%%%%%%%%%%%%%%%%%%%%

\section{Conclusions and Outlook\label{sec:conclusions}}

The CMRS format is designed specifically
for optimizing the SpMV operation
on modern graphics processing units.
It has several features distinguishing it from other formats developed for the same purpose:
(i)~it is an extension of a popular format, CRS, with a quick and
in-place conversion to and from it;
(ii) it has an efficient, single-kernel implementation;
(iii) it allows for dynamic assignment  of threads to the matrix rows;
(iv) it does not \emph{require} zero-padding,
row reordering nor any other matrix transformations for good performance;
(v) it has a great potential for off-line optimization techniques, including  zero-padding and row reordering,
which improve its efficiency for extremely  sparse matrices and
can turn it into one of the most efficient SpMV formats for GPUs.
Property (iii) distinguishes CMRS from other derivatives of the CRS format in which
a warp processes more than one matrix row, while properties (ii) and (v) distinguish it from the HYB
format.
These features should facilitate its adoption in existing software
and open new possibilities for its further optimization.
Moreover, the fact that our CMRS-based implementation of the SpMV kernel often approaches the hardware limit
suggests that this format will scale well into future GPU architectures.

The performance model of the CMRS SpMV kernel, despite its simplicity, turns out to fit
the actual results of numerical experiments well.
This indicates that the structure of typical sparse matrices from the UF SMC
is at least partially ordered---otherwise the SpMV efficiency would be determined
by indirect addressing of the input vector, a factor completely neglected in the model.
The model explains the acceleration of the CMRS over the standard vector kernel. It also
identifies the mean number of nonzero elements per row ($\mu$) as
a relevant parameter for the CRS-based SpMV kernels on GPUs.
The bandwidth efficiency $\eta_-$  (or $\eta_+$) can be used to
identify matrices for which further optimization is required as well as
help determine the quality of hardware support for the SpMV operation.

Our current implementations of the CMRS kernel are not without limitations.
The column index is stored on only 28 bits, which might prove insufficient
for future devices with larger amounts of memory.
However,
applications usually store much more data than just a single sparse matrix. For example,
a GPU-based computational fluid dynamics  solver may require $\approx 500$ bytes of storage per
each column of several of its sparse matrices \cite{Tomczak13}, which corresponds to
the memory threshold at $2^{28}\times 500$~B~$\approx 130$~GB, far above
the 12~GB available in modern accelerators.
Consequently, compression of the column index should not become a serious problem very soon.
A much more serious problem is related to the fact that CMRS is inherently limited by the amount of the
shared memory per multiprocessor. While in most cases this can be circumvented by
a suitable matrix transformation, as explained in Sec.~\ref{subsec:boost},
its efficient usage on Kepler-class GPUs for matrices
with values occupying more than 8 bytes, e.g., double precision complex numbers, may be problematic.
It is also not clear whether CMRS can be efficiently implemented on architectures lacking a
programmable, on-chip shared memory buffer. Moreover, CMRS requires the matrix to be sufficiently large.
Further research is also required to find the best ``off-line''
CMRS matrix optimization strategy---while our preliminary results with zero-padding
are very encouraging, our approach is rather complex and is not universal.

Finally, our results reveal the importance of developing a representative collection (or collections)
of sparse matrices for which the SpMV product is a truly relevant operation.
The UF SMC contains some very unusual matrices
for which the SpMV product is unlikely to be applicable, e.g.\ matrices with empty rows or columns.
Such atypical matrices can obfuscate the general picture
of the SpMV performance and its dependence on the matrix format and techniques used to implement it.

%%%%%%%%%%%%%%%%%%%%%%%%%%%%%%%%%%%%%%%%%%%%%%%%%%%%%%%%%%%%
%%%%%%%%%%%%%%%%%%%%%%%%%%%%%%%%%%%%%%%%%%%%%%%%%%%%%%%%%%%%
%%%%%%%%%%%%%%%%%%%%%%% SECTION %%%%%%%%%%%%%%%%%%%%%%%%%%%%
%%%%%%%%%%%%%%%%%%%%%%%%%%%%%%%%%%%%%%%%%%%%%%%%%%%%%%%%%%%%
%%%%%%%%%%%%%%%%%%%%%%%%%%%%%%%%%%%%%%%%%%%%%%%%%%%%%%%%%%%%

\section*{Acknowledgments}

ZK and MM prepared this publication as part of the  project of the City of  Wroc{\l}aw, entitled  ``Green Transfer'' --
aca\-demia-to-business knowledge transfer project co-financed by the European Union under the European Social Fund, under
the Operational Programme Human Capital (OP HC): sub-measure 8.2.1. SS acknowledges support from
Polish Ministry of Science and Higher Education Grant No. N N519 437939.

%\bibliographystyle{siam}
%\bibliography{speedit}

\begin{thebibliography}{10}

\bibitem{Bell08}
{\sc N.~Bell and M.~Garland}, {\em Efficient sparse matrix-vector
  multiplication on {CUDA}}, NVIDIA Technical Report NVR-2008-004, NVIDIA
  Corporation, Dec. 2008.

\bibitem{Bell09}
\leavevmode\vrule height 2pt depth -1.6pt width 23pt, {\em Implementing sparse
  matrix-vector multiplication on throughput-oriented processors}, in {SC}'09:
  Proceedings of the Conference on High Performance Computing Networking,
  Storage and Analysis, New York, NY, USA, 2009, ACM, pp.~1--11.

\bibitem{CUSP}
\leavevmode\vrule height 2pt depth -1.6pt width 23pt, {\em {CUSP}: A {C}++
  templated sparse matrix library}, 2012.
\newblock Version 0.3.0.

\bibitem{Baskaran2009}
{\sc R.~Bordawekar and M.~M. Baskaran}, {\em Optimizing sparse matrix-vector
  multiplication on {GPUs}}, Tech. Report RC24704, IMB Research, April 2008.

\bibitem{Choi:2010}
{\sc J.~W. Choi, A.~Singh, and R.~W. Vuduc}, {\em Model-driven autotuning of
  sparse matrix-vector multiply on gpus}, SIGPLAN Not., 45 (2010),
  pp.~115--126.

\bibitem{Florida}
{\sc T.~A. Davis and Y.~Hu}, {\em The university of {F}lorida sparse matrix
  collection}, ACM Trans. Math. Softw., 38 (2011), pp.~1:1--1:25.

\bibitem{DuBois:AnI2008}
{\sc D.~DuBois, A.~DuBois, T.~Boorman, C.~Connor, and S.~Poole}, {\em {An
  Implementation of the Conjugate Gradient Algorithm on FPGAs}}, in FCCM '08:
  Proceedings of the 2008 16th International Symposium on Field-Programmable
  Custom Computing Machines, Washington, DC, USA, 2008, IEEE Computer Society,
  pp.~296--297.

\bibitem{DuBois:Spa2008}
{\sc D.~DuBois, A.~DuBois, C.~Connor, and S.~Poole}, {\em Sparse matrix-vector
  multiplication on a reconfigurable supercomputer}, Field-Programmable Custom
  Computing Machines, Annual IEEE Symposium on,  (2008), pp.~239--247.

\bibitem{Dziekonski11}
{\sc A.~Dziekonski, A.~Lamecki, and M.~Mrozowski}, {\em A memory efficient and
  fast sparse matrix vector product on a {GPU}}, Progress {in} Electromagnetics
  Research, 116 (2011), pp.~49--63.

\bibitem{Farber2011}
{\sc R.~Farber}, {\em CUDA Application Design and Development}, Morgan
  Kaufmann, 1~ed., 2011.

\bibitem{Feng2011}
{\sc X.~Feng, H.~Jin, R.~Zheng, K.~Hu, J.~Zeng, and Z.~Shao}, {\em Optimization
  of sparse matrix-vector multiplication with variant {CSR} on {GPU}s}, in
  Parallel and Distributed Systems (ICPADS), 2011 IEEE 17th International
  Conference on, IEEE, 2011, pp.~165--172.

\bibitem{Garland:Spa2008}
{\sc M.~Garland}, {\em Sparse matrix computations on manycore {GPUs}}, in DAC
  '08: Proceedings of the 45th annual conference on Design automation, New
  York, NY, USA, 2008, ACM, pp.~2--6.

\bibitem{Grewe2011}
{\sc D.~Grewe and A.~Lokhmotov}, {\em Automatically generating and tuning {GPU}
  code for sparse matrix-vector multiplication from a high-level
  representation}, in GPGPU-4, Proceedings of the Fourth Workshop on General
  Purpose Processing on Graphics Pro-cessing Units ({GPGPU2011}), ACM, 2011,
  p.~12.

\bibitem{Liu:2013}
{\sc X.~Liu, M.~Smelyanskiy, E.~Chow, and P.~Dubey}, {\em Efficient sparse
  matrix-vector multiplication on x86-based many-core processors}, in
  Proceedings of the 27th international ACM conference on International
  conference on supercomputing, ICS '13, New York, NY, USA, 2013, ACM,
  pp.~273--282.

\bibitem{Maggioni13}
{\sc M.~Maggioni and T.~Berger-Wolf}, {\em An architecture-aware technique for
  optimizing sparse matrix-vector multiplication on {GPUs}}, Procedia Computer
  Science, 18 (2013), pp.~329 -- 338.

\bibitem{Malecha11}
{\sc Z.~Malecha, {\L}.~Miros{\l}aw, T.~Tomczak, Z.~Koza, M.~Matyka,
  W.~Tarnawski, and D.~Szczerba}, {\em {GPU-based simulation of 3D blood flow
  in abdominal aorta using OpenFOAM}}, Arch. Mech., 63 (2011), pp.~137--161.

\bibitem{Monakov10}
{\sc A.~Monakov, A.~Lokhmotov, and A.~Avetisyan}, {\em Automatically tuning
  sparse matrix-vector multiplication for {GPU} architectures}, in High
  Performance Embedded Architectures and Compilers, Y.~Patt, P.~Foglia,
  E.~Duesterwald, P.~Faraboschi, and X.~Martorell, eds., vol.~5952 of Lecture
  Notes in Comput. Sci., Springer Berlin / Heidelberg, 2010, pp.~111--125.
\newblock 10.1007/978-3-642-11515-8\_10.

\bibitem{Mukunoki2013}
{\sc D.~Mukunoki and D.~Takahashi}, {\em Optimization of sparse matrix-vector
  multiplication for {CRS} format on {NVIDIA} {K}epler architecture {GPU}s}, in
  Computational Science and Its Applications -- ICCSA 2013, B.~Murgante,
  S.~Misra, M.~Carlini, C.~M. Torre, H.-Q. Nguyen, D.~Taniar, B.~O. Apduhan,
  and O.~Gervasi, eds., vol.~7975 of Lecture Notes in Computer Science,
  Springer Berlin Heidelberg, 2013, pp.~211--223.

\bibitem{CUDA55}
{\sc NVIDIA}, {\em {CUDA C Programming Guide Version 5.5}}, May 2013.

\bibitem{Saule:arxiv:2013}
{\sc E.~Saule, K.~Kaya, and {\"U}.~V. \c{C}ataly{\"u}rek}, {\em Performance
  evaluation of sparse matrix multiplication kernels on {I}ntel {X}eon {P}hi}.
\newblock arXiv:1302.1078 [cs.PF], 2013.

\bibitem{Tomczak13}
{\sc T.~Tomczak, K.~Zadarnowska, Z.~Koza, M.~Matyka, and {\L}.~Miros{\l}aw},
  {\em Acceleration of iterative {N}avier-{S}tokes solvers on graphics
  processing units}, Int. J. Comput. Fluid Dyn., 27 (2013), pp.~201--209.

\bibitem{Vazquez2012}
{\sc F.~V\'azquez, J.~J. Fern\'andez, and E.~M. Garz\'on}, {\em Automatic
  tuning of the sparse matrix vector product on {GPU}s based on the {ELLR-T}
  approach}, Parallel Comput., 38 (2012), pp.~408--420.

\bibitem{Vazquez2009}
{\sc F.~V{\'a}zquez, E.~M. Garz{\'o}n, J.~A. Mart{\i}nez, and J.J.
  Fern{\'a}ndez}, {\em {Accelerating sparse matrix vector product with GPUs}},
  in Proceedings of the International Conference on Computational and
  Mathematical Methods in Science and Engineering (CMMSE 2009), CMMSE, 2009,
  pp.~1081--1092.

\bibitem{Vazquez2010}
{\sc F.~V{\'a}zquez, G.~Ortega, J.J. Fern{\'a}ndez, and E.M. Garz{\'o}n}, {\em
  Improving the performance of the sparse matrix vector product with {GPUs}},
  in 2010 10th IEEE International Conference on Computer and Information
  Technology (CIT 2010), IEEE Computer Society, 2010, pp.~1146--1151.

\bibitem{Vuduc:OSKI}
{\sc R.~Vuduc, J.~W. Demmel, and K.~A. Yelick}, {\em Oski: A library of
  automatically tuned sparse matrix kernels}, Journal of Physics: Conference
  Series, 16 (2005), p.~521.

\bibitem{Vuduc:Aut2003}
{\sc R.~W. Vuduc}, {\em Automatic performance tuning of sparse matrix kernels},
  PhD thesis, University of California, Berkeley, 2003.

\bibitem{Williams:Opt2007}
{\sc S.~Williams, L.~Oliker, R.~Vuduc, J.~Shalf, K.~Yelick, and J.~Demmel},
  {\em Optimization of sparse matrix-vector multiplication on emerging
  multicore platforms}, in SC '07: Proceedings of the 2007 ACM/IEEE conference
  on Supercomputing, New York, NY, USA, 2007, ACM, pp.~1--12.

\bibitem{Yang11}
{\sc X.~Yang, S.~Parthasarathy, and P.~Sadayappan}, {\em Fast sparse
  matrix-vector multiplication on {GPU}s: Implications for graph mining},
  Proceedings of the VLDB Endowment, 4 (2011), pp.~231--242.

\bibitem{Yoshizawa2012}
{\sc H.~Yoshizawa and D.~Takahashi}, {\em Automatic tuning of sparse
  matrix-vector multiplication for {CRS} format on {GPU}s}, 2012 IEEE 15th
  International Conference on Computational Science and Engineering, 0 (2012),
  pp.~130--136.

\end{thebibliography}

\clearpage

\centerline{\Large Additional Material}

\section{Derivation of Eq.~(5.1)}

\begin{figure}[h]
\begin{center}
  \includegraphics[width=0.5\columnwidth]{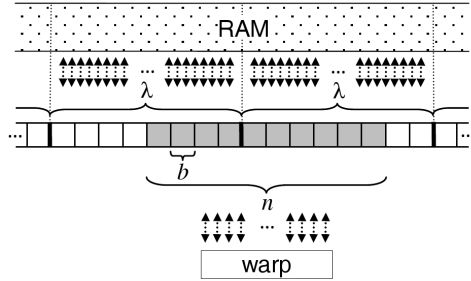}
  \caption{A schematic of the GPU memory access model.
  If a warp attempts to access a chunk of $n$ consecutive bytes made up of $b$-byte long words, this
  request is serviced by the L2 cache (middle), which is serviced by the main memory
  only through $\lambda$-byte-long,  $\lambda$-byte-aligned memory segments.
  If the requested chunk is located randomly relative to the segment boundaries,
  such an access pattern leads to the bandwidth waste and low kernel efficiency, unless $n \gg \lambda$.
  }\label{fig:add-model_explained}
\end{center}
\end{figure}

Let $n$ denote the size of the memory chunk (in bytes) to be accessed by a device that communicates with the
global memory only through aligned memory segments of $\lambda$ bytes. For the SpMV kernel
the value of $n$ is a multiple of $b=4,8$
(the size of the data items stored in the chunk)
and $\lambda $ is a multiple of 32 on modern GPUs (Nvidia suggests $\lambda=128$ for $b=4$).
Moreover, the chunk is aligned to $b$ bytes
(see Fig.~\ref{fig:add-model_explained}).
Let also assume that the number of memory transactions necessary to access the memory chunk is equal to
the minimum number of segments covering it.

Let $n = x\lambda - yb$,
where $x,y$ are integers,  $0 < x, 0 \le y < \lambda/b$.
If the beginning of the $b$-byte-aligned chunk is located
randomly relative to the memory segment boundaries,
then the number of distinct memory segments is  $x$ with probability $(y + 1)/(\lambda/b)$ and $x+1$ with probability
$1 - (y+1)/(\lambda/b)$.
Thus, the expected  number of memory segments accessed by the chunk is
$$
   S = x\frac{y+1}{\lambda/b} +  (x + 1)\left(1 - \frac{y+1}{\lambda/b}\right) = 1 + \frac{n - b}{\lambda}.
$$
Consequently, the ratio of the bytes transferred, $\lambda S$, to the bytes actually requested, $n$, is
$$
   \frac{\lambda S}{n} = \frac{\lambda + n - b}{n} =  1 + \frac{\lambda - b}{n}.
$$
Substituting $n = h\mu b$, one arrives at (\ref{eq:def-f}).

Note that the SpMV kernel actually transfers several independent streams of data, but for each of them Eq.~(5.1) predicts the same bandwidth efficiency. This justifies the usage of this equation for the
SpMV kernel.

\newpage

\section{Basic CMRS Kernel}
~~

\begin{lstlisting}[language=c++,
                   basicstyle=\footnotesize\sffamily,
                   commentstyle=\itshape,
                   columns=fullflexible]
// The basic CRS kernel is implemented as a a device function
// to facilitate experiments with dynamically vs. statically allocated shared memory
template<int HEIGHT, typename T,  bool USE_TEXTURE, bool ALIGN_DATA>
__device__
inline void
device_cmrs_multiply_original (
  const T* const __restrict__ X,  // input vector
  const int*     const __restrict__ stripe_offset,
  const int*     const __restrict__ col_idx,
  const T* const __restrict__ A,  // matrix values
  T *      const __restrict__ R,  // result vector
  unsigned       const num_rows,
  T   volatile * ptr  // pointer to shared memory
)
{
  const int thread_id   = blockDim.x * blockIdx.x + threadIdx.x;
  const int warp_id     = thread_id / WARP_SIZE;
  const int thread_lane = threadIdx.x & (WARP_SIZE-1);
  const int num_warps   = ( (blockDim.x + WARP_SIZE - 1) / WARP_SIZE) * gridDim.x;

  // a warp can process several strips to balance their sizes
  for(int stripe = warp_id; stripe*HEIGHT < num_rows; stripe += num_warps)
  {
    for(int k = 0; k < HEIGHT; k++)
      ptr[thread_lane + WARP_SIZE*k] = 0;

    const int stripe_start = stripe_offset[stripe];
    const int stripe_end   = stripe_offset[stripe + 1];
    // stripe_mid is used only if ALIGN_DATA == true
    const int stripe_mid   = ALIGN_DATA ?
        min(stripe_end, stripe_start - (stripe_start & 31) + 32) : stripe_start;

    // this attempts to read unaligned portion of the strip
    if (ALIGN_DATA)
    {
      int j = stripe_start + thread_lane;
      if (j < stripe_mid)
      {
        int c = col_idx[j];
        int r = c % CMRS_MAX_HEIGHT; // We use CMRS_MAX_HEIGHT == 16
        c >>= CMRS_BITS;  // We use CMRS_BITS == 4
        // macro fetch_x reads from an array directly or via one of the caches
        T xx = fetch_x<USE_TEXTURE>(c, X); // xx = X[c];
        xx *= A[j];
        r += HEIGHT * thread_lane;
        ptr[r] += xx;
      }
    }

    // standard CMRS loop
    for(int j = stripe_mid + thread_lane; j < stripe_end; j += WARP_SIZE)
    {
      int c = col_idx[j];
      int r = c % CMRS_MAX_HEIGHT; // We use CMRS_MAX_HEIGHT = 16
      c >>= CMRS_BITS; // We use CMRS_BITS == 4
      T xx = fetch_x<USE_TEXTURE>(c, X); // xx = X[c];
      xx *= A[j];
      r += HEIGHT * thread_lane;
      ptr[r] += xx;
    }

    // Now the parallel reduction for arbitrary 1 <= HEIGHT <= 16
    // We assume WARP_SIZE == 32

    T z = ptr[thread_lane]; // not sure if this register helps...

// #1
    ptr[thread_lane]   += ptr[thread_lane + HEIGHT*16];
    if (HEIGHT >= 4 or (HEIGHT == 3 && thread_lane < 16) )
       ptr[thread_lane + 1*32]   += ptr[thread_lane + HEIGHT*16 + 1*32];
    if (HEIGHT >= 6 or (HEIGHT == 5 && thread_lane < 16) )
      ptr[thread_lane + 2*32]   += ptr[thread_lane + HEIGHT*16 + 2*32];
    if (HEIGHT >= 8 or (HEIGHT == 7 && thread_lane < 16) )
      ptr[thread_lane + 3*32]   += ptr[thread_lane + HEIGHT*16 + 3*32];
    if (HEIGHT >= 10 or (HEIGHT == 9 && thread_lane < 16) )
      ptr[thread_lane + 4*32]   += ptr[thread_lane + HEIGHT*16 + 4*32];
    if (HEIGHT >= 12 or (HEIGHT == 11 && thread_lane < 16) )
      ptr[thread_lane + 5*32]   += ptr[thread_lane + HEIGHT*16 + 5*32];
    if (HEIGHT >= 14 or (HEIGHT == 13 && thread_lane < 16) )
      ptr[thread_lane + 6*32]   += ptr[thread_lane + HEIGHT*16 + 6*32];
    if (HEIGHT >= 16 or (HEIGHT == 15 && thread_lane < 16) )
      ptr[thread_lane + 7*32]   += ptr[thread_lane + HEIGHT*16 + 7*32];

 // #2
    ptr[thread_lane]          += ptr[thread_lane + HEIGHT*8];
    if (HEIGHT >= 5)
      ptr[thread_lane + 1*32]   += ptr[thread_lane + HEIGHT*8 + 1*32];
    if (HEIGHT >= 9)
      ptr[thread_lane + 2*32]   += ptr[thread_lane + HEIGHT*8 + 2*32];
    if (HEIGHT >= 13)
      ptr[thread_lane + 3*32]   += ptr[thread_lane + HEIGHT*8 + 3*32];

// #3
    ptr[thread_lane]   += ptr[thread_lane + HEIGHT*4];
    if (HEIGHT >= 9)
      ptr[thread_lane + 1*32]   += ptr[thread_lane + HEIGHT*4 + 1*32];

// #4
     z =  ptr[thread_lane]   += ptr[thread_lane + HEIGHT*2];

// #5
     z  += ptr[thread_lane + HEIGHT];

    // write the results
    int row = stripe*HEIGHT + thread_lane;
    if (thread_lane < HEIGHT && row < num_rows)
    {
      R[row] = z;
    }
  }
}
\end{lstlisting}

\newpage

\section{Modified CMRS Kernel}
~~

\begin{lstlisting}[language=c++,
                   basicstyle=\footnotesize\sffamily,
                   commentstyle=\itshape,
                   columns=fullflexible]
template<int MODULO, bool USE_TEXTURE, typename T>
__global__ void
cmrs_multiply(
  const T   * const __restrict   X,    // input vector
  const int * const __restrict__ stripe_offset,  // strip offset, see the paper
  const int * const __restrict__ col_idx,  // contains ColInd AND RowInStrip arrays, see the paper
  const   T * const __restrict__ A,    // matrix values
  T * const __restrict__ R,    // result vector
  unsigned  const              num_rows)
{
  const int HEIGHT = 16;  // fixed strip height
  const int asize = HEIGHT*MODULO;  // size of warp-owned array in shared memory

  extern __shared__ char cdata[];   // shared memory is assigned dynamically at kernel invocation

  // let's pretend the buffer contains T's
  T volatile * sdata = reinterpret_cast<T volatile *>(cdata);

  // let ptr point to the warp-owned buffer in shared memory; ptr = sdata[warp_lane];
  T volatile * ptr = &sdata[(threadIdx.x / WARP_SIZE)*asize];

  const int thread_id   = blockDim.x * blockIdx.x + threadIdx.x;
  const int warp_id     = thread_id / WARP_SIZE;
  const int thread_lane = threadIdx.x % WARP_SIZE;
  const int num_warps   = ( (blockDim.x + WARP_SIZE - 1) / WARP_SIZE) * gridDim.x;

  for(int stripe = warp_id; stripe*HEIGHT < num_rows; stripe += num_warps)
  {
    // let's zero the local buffer
    if (MODULO > 1)
    {
#pragma unroll
      for(int k = 0; k < MODULO/2; k++)
      {
        ptr[thread_lane + WARP_SIZE*k] = 0;
      }
    }
    else
    {
       if (thread_lane < HEIGHT)
         ptr[thread_lane] = 0;
    }

    // see the paper for what is going on here
    const int stripe_start = stripe_offset[stripe];
    const int stripe_end   = stripe_offset[stripe + 1];

    for(int j = stripe_start + thread_lane; j < stripe_end; j += WARP_SIZE)
    {
      int c = col_idx[j];
      int r = c % CMRS_MAX_HEIGHT;  // We use CMRS_MAX_HEIGHT == 16 == 2**4
      c >>= CMRS_BITS;              // We use CMRS_BITS == 4

      T xx = fetch_x<USE_TEXTURE>(c, X); // xx = X[c];
      xx *= A[j];

      r +=  HEIGHT*(thread_lane % MODULO); // modulo op. is the essence of the modified kernel
      ptr[r] += xx;
    }

    // Now the parallel reduction of the data pointed by ptr.
    // The size of the array pointed by ptr depends on MODULO, hence many conditionals.
    // In the paper we use MODULO = 8, sometimes 4 and 2.
    // MODULO = 1, 16 and 32 were used in tests and are also supported below.
    // Caveat: NVIDIA discourages the coding style that neglects __syncthreads()
    //   and relies on implicit inter-warp thread synchronization in future architectures.

    T z = 0; // not sure if this register helps...

    if (MODULO == 2 || (MODULO  == 1 && thread_lane < HEIGHT))
      z = ptr[thread_lane];

    if (MODULO == 32)
    {
      ptr[thread_lane]        += ptr[thread_lane + HEIGHT*16];
      ptr[thread_lane + 1*32] += ptr[thread_lane + HEIGHT*16 + 1*32];
      ptr[thread_lane + 2*32] += ptr[thread_lane + HEIGHT*16 + 2*32];
      ptr[thread_lane + 3*32] += ptr[thread_lane + HEIGHT*16 + 3*32];
      ptr[thread_lane + 4*32] += ptr[thread_lane + HEIGHT*16 + 4*32];
      ptr[thread_lane + 5*32] += ptr[thread_lane + HEIGHT*16 + 5*32];
      ptr[thread_lane + 6*32] += ptr[thread_lane + HEIGHT*16 + 6*32];
      ptr[thread_lane + 7*32] += ptr[thread_lane + HEIGHT*16 + 7*32];
    }

    if (MODULO > 8)
    {
      ptr[thread_lane]        += ptr[thread_lane + HEIGHT*8];
      ptr[thread_lane + 1*32] += ptr[thread_lane + HEIGHT*8 + 1*32];
      ptr[thread_lane + 2*32] += ptr[thread_lane + HEIGHT*8 + 2*32];
      ptr[thread_lane + 3*32] += ptr[thread_lane + HEIGHT*8 + 3*32];
    }

    // here starts the parallel reduction for MODULO==8, as used in the paper
    if (MODULO > 4)
    {
      ptr[thread_lane]        += ptr[thread_lane + HEIGHT*4];
      ptr[thread_lane + 1*32] += ptr[thread_lane + HEIGHT*4 + 1*32];
    }

    if (MODULO > 2)
    {
      z =  ptr[thread_lane]   += ptr[thread_lane + HEIGHT*2];
    }

    if (thread_lane < HEIGHT && MODULO > 1)
    {
      z  += ptr[thread_lane + HEIGHT];
    }

    // writing the results to R
    int row = stripe*HEIGHT + thread_lane;
    if (thread_lane < HEIGHT && row < num_rows)
    {
      R[row] = z;
    }
  }
}
\end{lstlisting}

\end{document}